\documentclass[sigconf,balance=false]{acmart}
\usepackage{popets}
\usepackage{graphicx}
\usepackage{tabularx}
\usepackage{tikz}
\usepackage{amsmath}
\usepackage{filecontents}
\usepackage{booktabs}
\usepackage{array}
\usepackage{caption}
\usepackage{caption}
\usepackage{subcaption}
\usepackage{pgfplots}
\usepackage{hyperref}
\usepackage{multirow}

\setcopyright{none}
\renewcommand\footnotetextcopyrightpermission[1]{} 
\acmYear{}
\acmVolume{}
\acmNumber{}

\acmConference{}
\newcounter{rcounter}
\newenvironment{rec}[1][]{\refstepcounter{rcounter}\par\itshape\medskip
   \noindent\textit{\textbf{Recommendation \thercounter:} #1}}{\medskip}

\begin{document}
\startPage{1}

\title{Two Steps Forward and One Step Back: The Right to Opt-out of Sale under CPRA}


\author{Jan Charatan}
\affiliation{%
  \institution{Pomona College}
  \city{}
  \state{}
  \country{}}
\email{jbca2019@mymail.pomona.edu}

\author{Eleanor Birrell}
\affiliation{%
  \institution{Pomona College}
  \city{}
  \state{}
  \country{}
  }
\email{eleanor.birrell@pomona.edu}


\renewcommand{\shortauthors}{Charatan and Birrell}

\begin{abstract}
The California Privacy Rights Act (CPRA) was a ballot initiative that 
revised the California Consumer Privacy Act (CCPA). Although often framed as expanding 
and enhancing privacy rights, a close 
analysis of textual revisions---both changes from the earlier law and changes
from earlier drafts of the CPRA guidelines---suggest that the reality might be
more nuanced. In this work, we identify three textual revisions that have potential
to negatively impact the right to opt-out of sale under CPRA and evaluate the 
effect of these textual revisions using (1) a large-scale longitudinal measurement
study of 25,000 websites over twelve months and (2) an experimental user study with 
775 participants recruited through Prolific. We find that all revisions negatively
impacted the usability, scope, and visibility of the right to opt-out of sale. 
Our results provide the first comprehensive evaluation of the impact of CPRA on
Internet privacy. They also emphasize the importance of continued evaluation of
legal requirements as guidelines and case law evolve after a law goes into effect. 
\end{abstract}


\maketitle
\pagestyle{plain}

\section{Introduction}

Online activity---including which websites people visit and how people interact with those sites---can be used to accurately predict information about interests and demographics~\cite{hinds2018demographic}. As a result, this information is a valuable commodity, and many websites sell information collected about online activity to third parties, including advertisers and data analytics companies. This type of data collection and sale runs contrary to many users' preferences~\cite{o2021clear}. However, legal efforts to curtail this practice, such as the right to opt-out of sale of personal information introduced by the California Consumer Privacy Act (CCPA), have been ineffective because the resulting opt-out mechanisms are hard for users to find~\cite{o2021clear,3websitestudy} and exhibit deceptive design patterns that nudge users away from invoking their rights~\cite{o2021clear}.

The California Privacy Rights Act (CPRA), which went into effect in January 2023, is an amendment to CCPA. Although CPRA is often framed as expanding privacy rights for California consumers compared the earlier CCPA, the reality is more nuanced. Although CPRA grants new rights---such as a right to limit the use of sensitive personal information---and codifies earlier guidelines---such as explicitly detailing how automated opt-out signals must be handled and defining (and prohibiting) the use of dark patterns---the final CPRA guidelines also include provisions that might undermine earlier privacy rights such as the right to opt-out of sale. 

In this work, we identify three textual revisions introduced by CPRA that have the potential to weaken the right to opt-out of sale:
\begin{enumerate}
    \item \textbf{Reduced mechanisms:} CCPA required businesses to provide a opt-out link on their home page entitled ``Do Not Sell My Personal Information'' (Cal. Civ. Code \S 1798.135(a)(1)) and the California Privacy Protection Agency guidelines state that automated browser signals must be treated as valid opt-outs~\cite{cppafaqs}. CPRA exempts businesses that respect automated browser signals from the requirement to provide that manual opt-out link (Cal. Civ. Code \S 1798.135(b)(1)). 
    \item \textbf{Reduced scope:} CCPA applied to businesses that  buy, sell, or share the personal information of 50,000 or more consumers or households. CPRA increased this threshold to 100,000  (Cal. Civ. Code \S 1798.140(d)(1)(B)). 
    \item \textbf{Reduced visibility:} The first draft of the CPRA guidelines stated that when a business detects an opt-out signal, it ``should display whether or not it has processed the consumer’s opt-out preference signal''. The final guidelines stated that businesses ``may display'' such a notice (Cal. Civ. Code \S 7025(c)(6)). 
\end{enumerate}
We evaluate the effect of these three textual revisions through a combination of a longitudinal measurement study and an experimental user study. 

The textual revisions that reduce legally mandated opt-out mechanism and that reduce the scope of the California privacy law share the property that both the earlier text and the revised text were legally enforced under California law at different points in time. To evaluate the impact of these textual revisions on the right to opt-out of sale, we conducted a longitudinal measurement study over twelve months: we started collecting data in November 2022---two months before CPRA went into effect---and continued until November 2023---five months after enforcement of CPRA. Each month we visited 25,000 websites and automatically classified whether each site provided an opt-out of sale link, respected automated opt-out signals, and discussed opt-out mechanisms in its privacy policy. We found that although the number of sites that respect opt-out signals has increased since CPRA went into effect, the number of sites that only respect automated signals has also increased. Since many browsers to not yet support automated signals, this potentially reduces  accessibility of the right to opt-out of sale. We also found that 24\% of websites that provided opt-out of sale links prior to CPRA no longer provide any mechanism to opt-out of sale; this confirms that the reduced scope of CPRA has also negatively impacted privacy. 

\begin{figure*}[t!]
\includegraphics[width=.95\textwidth]{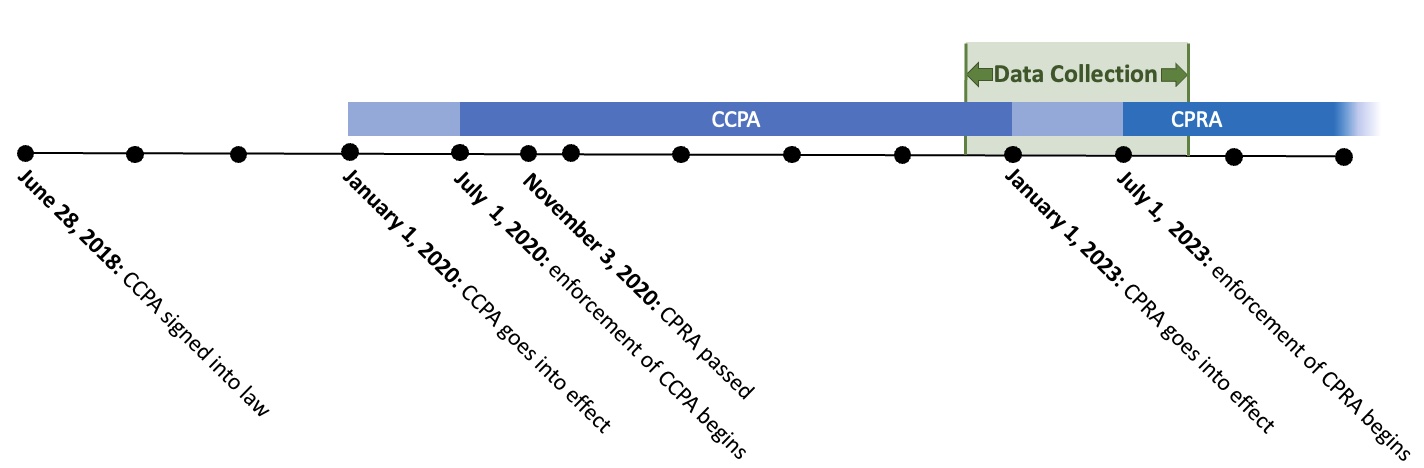}
\caption{Timeline of California privacy regulations.}\label{fig:timeline}
\end{figure*}

The textual revision that reduced visibility of how automated signals are handled was a revision to draft guidelines that were never legally in effect. To evaluate the impact of this revision, we conducted an experimental user study with 775 participant recruited through Prolific. Our user study measured the impact of various possible signal display designs compared to the baseline state where no display is provided. We found that eliminating these displays reduced awareness of and confidence about whether a website sells their personal information, although these effects were only significant if the display was sufficiently visible. These results suggest that reducing the visibility of how websites handle automated signals likely had a negative impact on consumer privacy, but that the efficacy of the earlier requirement would have depended on implementation guidelines and enforcement. 

Our results provide the first comprehensive evaluation of the impact of CPRA on
the right to opt-out of sale. They also emphasize the importance of continued evaluation of
legal requirements as guidelines and case law evolve after a law goes into effect.

\section{Background}

To provide context for our work, this section provides background on California privacy regulations and the technologies developed to implement the right to opt-out of sale under these laws. A timeline of the relevant regulations is shown in Figure~\ref{fig:timeline}. 

\paragraph{CCPA} The passage of the California Consumer Privacy Act of 2018 (CCPA)~\cite{ccpa} made California the first state in the United States to sign a comprehensive piece of consumer privacy legislation into law. CCPA granted California consumers four key rights: (1) the right to know about the data being collected about you, (2) the right to delete data about you, (3) the right to opt-out of the sale of your personal information, and (4) the right to non-discrimination for invoking CCPA rights. Under CCPA, both ``personal information'' and ``sale'' were broadly defined. ``Personal information'' included any information ``that identifies, relates to, describes, is reasonably capable of being associated with, or could reasonably be linked, directly or indirectly, with a particular consumer or household'', a  definition that explicitly included information about online activities (e.g., a user's interactions with a website) and any inferences drawn from personal information. ``Sale'' included both monetary transactions and any ``other valuable consideration''. CCPA applied to business that collected personal information from more than 50,000 consumers or that derived the majority of their profits from selling consumer data; non-profit organizations were exempt. CCPA went into effect on January 1, 2020 and enforcement began on July 1, 2020. 

The text of CCPA stated that companies must implement the right to opt-out of sale by providing a ``clear and conspicuous'' link entitled ``Do Not Sell My Personal Information'' on the homepage of their website that allowed users to invoke this right. After enforcement began in July 2020, the California Attorney General's office brought enforcement actions against companies whose mechanisms ``included choices that were confusing with unclear language and toggle option'' or ``directed consumers to a confusing webpage that required several additional steps to submit CCPA requests''. The AG's office also clarified that automated browser signals such as Global Privacy Control (GPC) signals must be treated as valid requests to opt-out of sale~\cite{gpctweet,cppafaqs} and subsequently brought enforcement actions against companies that ignored GPC signals~\cite{violation,ccpaenf}. 

\paragraph{CPRA} The California Privacy Rights Act (CPRA)~\cite{cpra} was a ballot proposition approved by a majority of California voters in November 2020. CPRA revised and amended CCPA in a variety of ways, including introducing two new rights---the right to correct inaccurate personal information and the right to limit the use and disclosure of sensitive personal information---and codifying requirements relating to dark patterns and automated signals. It also created a California Privacy Protection Agency (CPPA), which was tasked with enforcing the rights of Californians under the amended law, and it modified the scope of applicability to only include business that collect personal information from more than 100,000 consumers or that derive the majority of their profits from selling or sharing consumer data. Implementation details for CPRA were defined by the new CPPA: a first draft of these guidelines was released in July 2022~\cite{cpradraft0722} with updates released in November 2022~\cite{cpradraft1122} and March 2023~\cite{cprafinal}. CPRA went into effect on January 1, 2023 and enforcement began on July 1, 2023. 

\paragraph{Global Privacy Control (GPC)} GPC~\cite{gpc} is an opt-out signal that is sent via HTTP headers or the DOM. Unlike some proposed privacy preference signals, such as the Platform for Privacy Preferences (P3P)~\cite{p3p} or the Advanced Data Protection Control (ADPC)~\cite{adpc}, GPC is a simple, one-bit signal that requires no configuration or negotiation. Unlike signals such as Do Not Track (DNT) or the Network Advertising Initiative (NAI), companies are legally required to accept GPC as a valid mechanism to opt-out of sale under California law. GPC is currently implemented in some browsers---including Firefox, DuckDuckGo, and Brave---and is available as an extension for Chrome and Firefox. GPC is currently enabled by over 50 million users.

\paragraph{US Privacy Strings} To provide technical support for California's legal requirement that companies must respect GPC signals, the Interactive Advertising Bureau (IAB) Tech Lab introduced the US Privacy String~\cite{usprivacystring}, a specification and API for communicating signals about privacy choices under CCPA. US Privacy Strings are concise, four-character strings that communicate: (1) the version number, (2) whether the website has notified users about their data practices and provided an opportunity to opt-out, (3) whether the user has opted out of sale of their personal information under CCPA, and (4) whether the website has signed the IAB Limited Service Provider Agreement. The current version number is 1. For the remaining three fields, there are three possible values: Yes ("Y"), No ("N"), or CCPA/CPRA is not applicable ("-"). 

Websites are supposed to send copies of a US Privacy String along with any payload to enable third-parties to comply with the user's opt-out preferences. Upon receiving a GPC signal, a website must update its US Privacy String to indicate that the user has opted-out of sale. US Privacy Strings can be implemented as cookies or as JavaScript, and a website's US Privacy String can be accessed using the USP API.

\section{Related Work}

Although this work is the first to evaluate the impact of CPRA, prior work has previously evaluated CCPA's right to opt-out of sale and other rights under CCPA. Prior work has also studied opt-out signals in various contexts. 

\paragraph{Measuring Opt-out of sale under CCPA} Several projects have observed and measured how websites implement the right to opt-out of sale. A Consumer Reports study~\cite{3websitestudy} asked users to attempt to opt-out of sale on 216 websites from the California Data Broker registrar (all of whom sell data and were subject to CCPA); they found that 11.1\% of data broker websites lacked a CCPA-required Do Not Sell link on the homepage, and they documented examples of difficult, unclear, and time-intensive opt-out processes. O'Connor et al. manually evaluated 500 top websites longitudinally over the first year after enforcement of CCPA began~\cite{o2021clear}; although they found an increase in the prevalence of opt-out links, they documented many instances of websites that failed to provide legally-mandated opt-out links and found that manipulative and inconvenient designs were prevalent.  Later work by Van Nortwick and Wilson measured CCPA compliance rates by examining the state of do not sell links on almost 500,000 websites~\cite{van2022setting}; they found that only about 2\% of websites have such links and that only about 40\% of these links meet the minimum standards of readability. 

\paragraph{Evaluating CCPA Opt-outs} Some work has also evaluated the usability of opt-out mechanisms. The Consumer Reports study~\cite{3websitestudy} found found that 31.4\% of the sites studied displayed their link in such a manner that at least one out of three users was unable to find it, that more than a third of participants spent over five minutes opting out (with a maximum time of over an hour), and that 14\% were unable to successfully complete the process. O'Connor et al.~\cite{o2021clear} evaluated the effect of design patterns and nudging on opt-out rates; they found that the most common design---a link at the bottom of the page---resulted in an opt-out rate of 1.4\% compared to 12.2\% for designs with visible banners. They also found that manipulative designs and inconvenience factors significantly affect opt-out rates. 

\paragraph{Improving CCPA Opt-outs} Based on these problems with current opt-out mechanisms, some researchers have looked at ways that opting out could be made easier. Habib et al. proposed an icon that could help signal to users that they can exercise CCPA privacy choices \cite{habib2021toggles,cranor2020design}; their recommended design has since been incorporated into the legal guidelines for CCPA and CPRA. Zimmeck et al.~\cite{zimmeck2020standardizing} created a browser extension called OptMeOwt that automatically places Do Not Sell cookies on sites and sends Do Not Sell headers; this work has been expanded and standardized as the Global Privacy Control (GPC) signal. Siebel et al.~\cite{siebel2022impact} developed a browser extension that generated standardized, visible banners. Bannihatti et al.~\cite{bannihatti2020finding} developed a browser extension to find opt out statements in privacy policies using heuristics and machine learning.

\paragraph{Other CCPA Rights} There is also some prior work that focused on other rights under CCPA. Chen et al.~\cite{chen2021fighting} focused on the right to know and showed that even if websites are legally required under the CCPA to disclose certain data practices, variance and vagueness in definitions can lead to confusion among users. Samarin et al.~\cite{samarin2023lessons} issued access requests to mobile apps and compared the results to personal data observed through a analysis of app network traffic; they found that significant percentages of apps collected personal information that was not disclosed when users invoked their right to know. 

\paragraph{Opt-out Signals} Hils et al.~\cite{hils2021privacy} conducted a longitudinal study quantifying the adoption of privacy preference signals between 1997 and February 2021; however, their work looked at privacy signals more broadly and the timespan ended too soon to capture the effect of CCPA's revised guidelines that stated that GPC signals must be treated as valid opt-outs under CCPA. Human et al.~\cite{human2022data} conducted a technical comparison of different privacy preference signals; however, they didn't conduct any measurement studies or user studies. Zimmeck et al.~\cite{zimmeck2023usability} conducted a user survey that found that most people understood GPC and would enable it if GPC were supported by their browser. They also used a browser extension to analyze GPC compliance on 464 sites and found that only 12\% of sites respected GPC signals; that work was conducted in 2022 before CPRA went into effect.

\section{Measuring the Right to Opt-out of Sale Under CPRA}\label{sec:measurement_study}

The textual revisions that reduce legally mandated opt-out mechanism and that reduce the scope of the California privacy law share the property that both the earlier text and the revised text were legally enforced under California law at different points in time. To evaluate the impact of these textual revisions on the right to opt-out of sale, we conducted a longitudinal measurement study over twelve months---from November 2022 to November 2023---in which we visited 25,000 top sites and automatically classified whether each site provided an opt-out of sale link, respected automated opt-out signals, and discussed opt-out mechanisms in its privacy policy.

\subsection{Measurement Study Methodology}\label{sec:measurement_methods}

We implemented a webcrawler using Selenium~\cite{selenium} and GeckoDriver~\cite{geckodriver} and used this crawler to visit the top 25,000 websites from the October 29, 2022 Tranco top websites list~\cite{tranco}\footnote{We do not expect that all 25,000 websites on this list are subject to CCPA or CPRA. In fact, this dataset is known to contain non-English language sites and non-user facing domains. However, since we are trying to understand the impact of CPRA on the right to opt-out for California users---and not attempting to measure compliance rates with either regulation---we consider it more important to cast a wide net that represents the range of websites California users visit. Tranco's top-website list is well-represented in browser experience reports, is relatively stable over time, has reasonable responsiveness rates, and is large enough to include a wide range of frequently visited websites~\cite{lepochat2019evaluating}, making it an appropriate choice for this work.} thirteen times between mid-November 2022 and mid-November 2023. All of the data collection was done on AWS EC2 c6i.8xlarge instances that were located in a California data center (us-west-1) so that any CCPA functionality hidden behind geofencing would be shown to our crawler. 


For each website visited, our crawler logged the following:

\begin{enumerate}
    \item \textbf{Do Not Sell Links.} We recorded whether an opt-out of sale link was present on the website's homepage. This was determined by looking for phrases that are a slight variation of the legally-mandated text. We started with the list of phrasings used in prior work~\cite{van2022setting} and added additional language to include the new alternative opt-out link permitted under CPRA: ``Your Privacy Choices'' or ``Your California Privacy Choices''.
    \item \textbf{GPC Response.} To determine how websites respond to GPC signals, we visited each website twice: once with the GPC signal turned off and then a second time with the GPC signal turned on. During each visit, we used the USP API to access the website's US Privacy String (if provided). Websites whose US Privacy String indicated they had not received an opt-out signal during the first visit (when GPC was not enabled) and then indicated that they had received an opt-out signal during the second visit (when GPC was enabled) were considered to respect GPC signals. 
    \item \textbf{Privacy Policy Language.} To explore how website privacy policies talk about GPC, we used a heuristic approach to locate the website privacy policy: (1) We searched the homepage of the website for a link labeled ``Privacy Policy'' or ``Privacy Statement''. (2) If no such link was found, we conducted a secondary search for any link that contains the word ``privacy'' and then searched all the landing pages for links labeled ``Privacy Policy'' or ``Privacy Statement''. (3) If we failed to find a privacy policy link in both the primary and secondary phase, we classified the website as having no privacy policy accessible. If we found a privacy policy link in either the primary or secondary phase, we loading the link's destination page and searched it GPC-related terminology using three regular expressions: 
    
    \begin{itemize}
        \item \texttt{opt(-| )out( preference)? signals?( honored)?}
        \item \texttt{global privacy control}
	   \item \texttt{browser(-based( standard)?)? signals?}
    \end{itemize}

    \item \textbf{/.well-known/gpc.json} To be compliant with the GPC specification, websites are supposed to have a \texttt{gpc.json} file within the \texttt{/.well-known} directory of their website that contains a property \texttt{gpc} with the value \texttt{true}. For each website visited, we collected the \texttt{/.well-known/gpc.json} file, if such a file existed. 
\end{enumerate}

\paragraph{Data Consistency} During preliminary testing, we found that network delays and temporary website outages occasionally resulted in different runs collecting different data. To mitigate this effect, we ran our webcrawler twice for each sample collected and used the following approach for conflict resolution: 

\begin{itemize}
    \item If an opt-our of sale link was found on one of two runs, we report true for the overall run.
    \item For the US Privacy Strings and US API version, we first prefer complete data. If the runs are both complete and still different, we prefer the more recent run. If we have one run that only has the string and version for GPC off and another which only has the string and version for GPC on, then we stitch the runs together to produce a complete run.
    \item If we have conflicting data for privacy policies, we prefer the results in this order: privacy policy on home $\rightarrow$ privacy policy layer down $\rightarrow$ privacy link $\rightarrow$ no privacy link $\rightarrow$ error. If we have two searches that happened on the same layer and conflicting numbers, we prefer the run with more occurrences of regexes. 
    \item For the results of \texttt{GPC.json}, we prefer a run where the \texttt{GPC.json} is found. If there is a conflict between the contents, we prefer the more recent run.
\end{itemize}

\paragraph{Limitations} Although US Privacy Strings are a valuable indicator of how websites respond to GPC signals, they are an imperfect reflection of actual website behavior. Some websites that are subject to CCPA might not support US Privacy Strings, resulting in the omission of sites from our dataset. US Privacy Strings are also not set automatically; it would be possible for a website to detect GPC signals and update their privacy string without ceasing to sell or share a user's personal information. 

All data collection was conducted using the Chrome web browser. Since website behavior can vary between browsers, some users might experience different options when visiting sites from an alternate browser. 

Moreover, prior work has found that automated browser can be easily detected and that up to 14\% of websites include functionality to detect automated browsers~\cite{krumnow2022gullible}. In such cases, websites may behave differently when visited by an automated browser compared to when a human visits from a standard web browser. To quantify the impact of using an automated web browser to collect our data, we manually validated our results (Section~\ref{sec:validation}). 

\begin{figure}[t]
    \centering
    \includegraphics[width=\columnwidth]{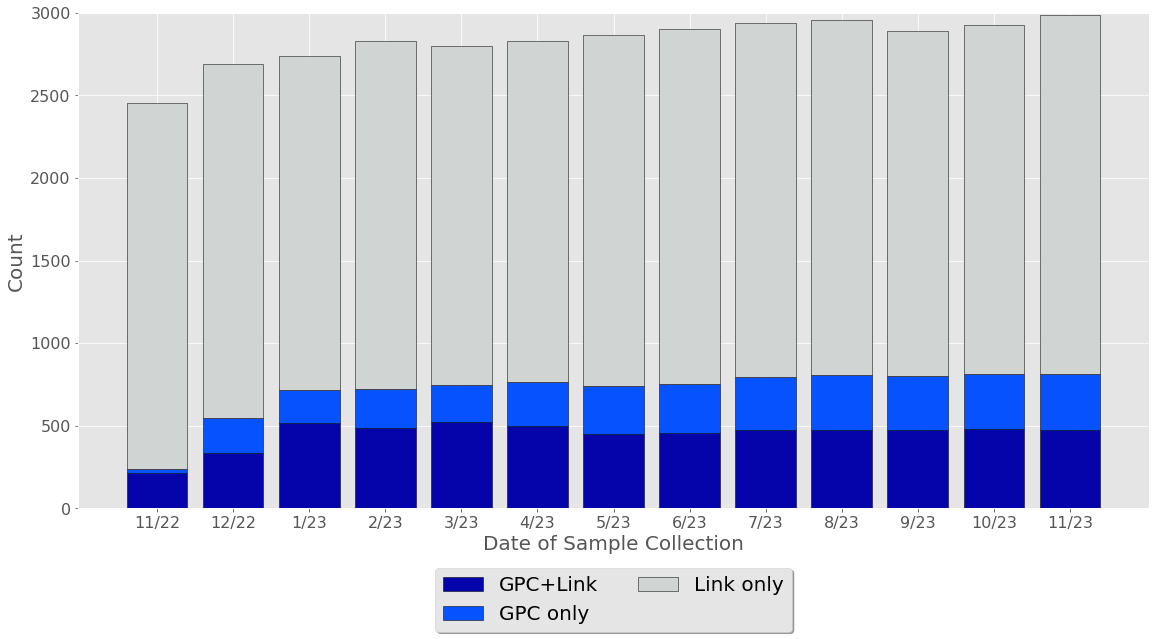}
    \caption{Frequency and type of opt-out of sale mechanisms provided by websites in the Tranco Top 25,000 dataset between November 2022 and November 2023.}
    \label{fig:optouts-frequencies}
\end{figure}

\subsection{Methodology Validation}\label{sec:validation}

After performing conflict resolution, we found that 1.7\% of websites visited exhibited some kind of discrepancy between the two runs. 0.1\% had an inconsistency in whether we detected an opt-out of sale link. 0.3\% had an inconsistency in what US Privacy String we detected. 1.4\% had inconsistencies when searching for GPC-related language in the privacy policy; we believe this higher rate is due to our crawler needing to load several different pages to complete that part of the data collection process, introducing additional opportunities for network delays to cause our script to time out. 

To validate our methodology further, we generated a striated sample of 100 websites by randomly selecting 10 websites from each range of 2,500 websites from the Tranco Top 25,000 list and manually replicated the data collection for each of those 100 sites. 

Our webcrawler was 96\% accurate at identifying the presence of an opt-out of sale link. Our manual replication discovered 3/100 false negatives (one website had the link implemented as a button, one had a link entitled ``Cookie Preferences / Do Not Sell'', and one had a link entitled ``Your California Privacy Rights'') and one false positive (on an international news website). We also validated our link detection by comparing directly to prior work: our crawler detected opt-out of sale links on 12.3\% of websites from the Tranco Top 10,000 in November 2022, a number that is consistent with  prior work that found links on 13.7\% on Tranco's Top 10,000 list from earlier in the year~\cite{rasaii2023exploring}. 

Our manual validation also showed that our webcrawler was 97\% accurate at identifying US Privacy Strings. On 2/100 websites, the crawler and the manual observation were inconsistent when GPC was not enabled (in both cases, the crawler received a privacy string that states that GPC was not enabled; one website's privacy string claimed it was detecting a GPC signal when visited manually from a browser without GPC enabled and the other claimed that CCPA was not applicable to the visitor). In 1/100 websites, the crawler identified a GPC string but no such strings were found when manually browsing the website. 

Our webcrawler was 89\% accurate at finding privacy policies. All failure cases were false negatives (i.e., the webcrawler failed to identify a privacy policy link that was detected through a manual search). Despite failing to locate some links, the webcrawler was 98\% accurate at recognizing whether a website's policy created GPC-specific language: on 2/100 websites the crawler failed to locate language because it had failed to find a link to that website's privacy policy. For all websites whose policies matched any of our GPC-language regular expressions, the crawler correctly identified which expressions were present in the policy, however in two instances the crawler counted one more occurrence that was rendered visibly in the browser. 

Our webcrawler was 100\% accurate at detecting whether websites in our manual sample provided \text{gpc.json} files.

While this manual validation shows that our crawler is not perfectly accurate, it also demonstrates that our crawler exhibits high accuracy overall. 

\begin{figure}[t]
    \centering
    \includegraphics[width=\columnwidth]{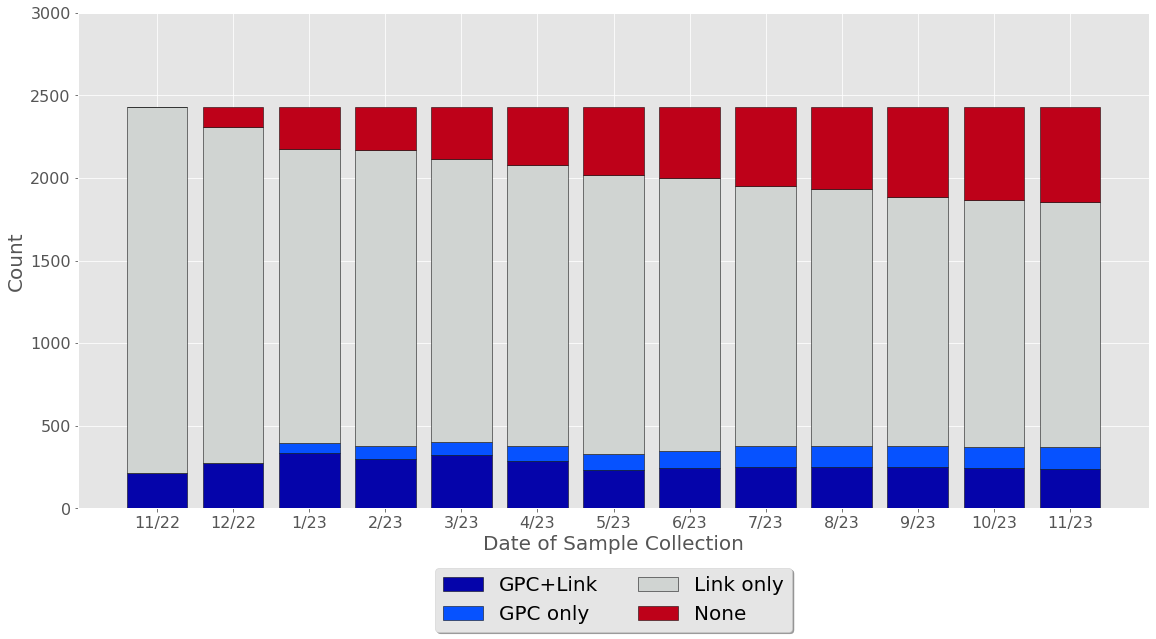}
    \caption{Opt-out of sale mechanisms provided by the 2,429 sites that had opt-out links in November 2022.}
    \label{fig:optouts}
\end{figure}

\subsection{Measurement Study Results}

To evaluate the impact of CPRA overall---and textual revisions specifically---on the right to opt-out of sale, we analyzed our results from multiple angles.

\subsubsection{Opt-out Mechanisms provided by CCPA-compliant sites}
\label{subsubsection:mechanisms}

We first measured the availability and type of opt-out of sale mechanism provided by websites in our dataset during the year between November 2022 and November 2023. These trends are shown in Figure~\ref{fig:optouts-frequencies}.

Of the top 25,000 sites, 2,452 provided some form of opt-out of sale mechanism in November 2022. 2,429 websites (9.9\%) provided manual opt-out of sale links on their homepage, as required at the time by CCPA; 212 of those sites also respected GPC in November 2022. An additional 23 websites respected GPC signals but did not provide a manual link. 

Over the next two months, we observed an increase in the number of websites that provided some form of opt-out of sale mechanism (to 2,737 in January 2023, shortly after CPRA went into effect). We also observed a sharp increase in compliance with GPC signals from 235 in November 2022 to 715 in January 2023. However, we also observed a sharp increase in the number of websites that respected GPC signals but provided no manual opt-out mechanism, from 23 in November 2022 to 197 in January 2023. Between January 2023 and November 2023, all three numbers continued to grow gradually. Our final crawl, conducted in November 2023, 2,987 websites provided some form of opt-out mechanism and 814 honored GPC signals. However, there were 340 websites for which GPC signals were the only supported mechanism for opting-out of sale.

While there are well-known usability issues with manual opt-out links---including the fact that many users struggle to find these links~\cite{3websitestudy} and that opt-out interfaces exhibit manipulative and inconvenient designs that nudge users away from invoking their right to opt-out~\cite{o2021clear}---the fact remains that some users do utilize these opt-out links and that most popular browsers do not yet support GPC signals. Thus while sites that support GPC only are legally compliant with CPRA, invoking the right to opt-out of sale on these sites is unavailable to most users. 

To further explore the effect of textual revisions on the availability of various types of opt-out mechanisms, we conducted additional analysis on the set of 2,429 websites that provided a manual opt-out of sale mechanism in November 2022. 
Although prior work indicates that this is likely a subset of the set of sites that were subject to CCPA~\cite{van2022setting}, focusing on this set allows us to quantify how opt-out mechanisms have changed among a set of websites that are trying to comply with California law.  
These results are summarized in Figure~\ref{fig:optouts}.

We found that prior to CPRA, few of these websites respected GPC signals: only 8.7\% respected GPC (i.e., updated their US Privacy String from N to Y when GPC was enabled) in November 2022, rising to 16.2\% in January 2023 when CPRA went into effect. This rate was stable in the 13.5-16.4\% range over the following ten months. 
However, at the same time we observed a decrease in the number of websites that provided manual opt-out links on their homepage. A small but growning subset of these sites switched from providing a manual opt-out of sale link to only accepting opt-outs sent as GPC signals. While these sites are legally compliant with CPRA, invoking the right to opt-out of sale on these sites is unavailable for most users. 
Most of the websites that provided opt-out of sale links in November 2022 and subsequently removed them provided no opt-out of sale mechanism after CPRA went into effect. Given that these websites previously complied with the requirements of CCPA, we believe that these websites represent the group that were subject to CCPA but are no longer subject to the revised scope because they collect personal data about more than 50,000 California consumers but fewer than 100,000 California consumers. This suggests that the reduced scope of CPRA reduced the set of companies subject to California privacy regulations by approximately 24\%, significantly decreasing the impact of this regulation.

\begin{figure}[t]
    \centering
    \includegraphics[width=\columnwidth]{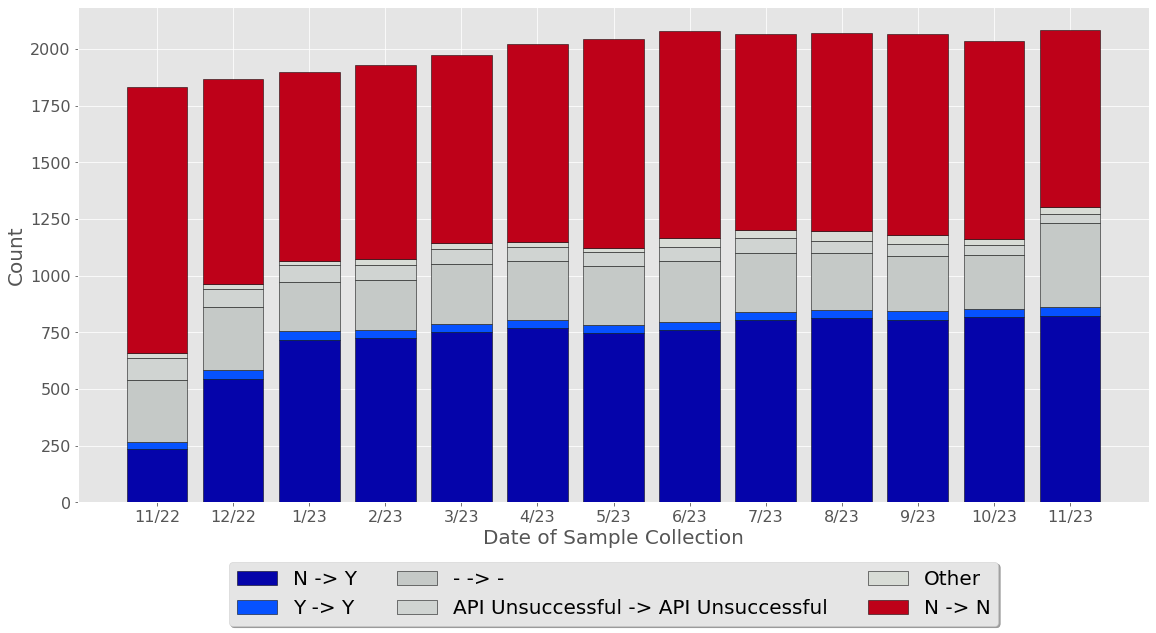}
    \caption{Changes in U.S. Privacy String when GPC is enabled, among the top 25,000 sites.}
    \label{fig:usps}
\end{figure}

\subsubsection{Compliance with GPC Opt-out Signals}

In addition to analyzing changes in opt-out of sale mechansisms, we also analyzed compliance with GPC signals among the top 25,000 websites more broadly. Our results are summarized in Figure~\ref{fig:usps}.

Most websites did not set a US Privacy String . While this number grew in the year between November 2022 and November 2023, the total number of websites that set any value of US Privacy String in November 2023 was only 8.3\%. This likely indicates that the other 91.7\% of websites ignore GPC signals entirely.

In our first data collection, conducted in November 2022, we found that just 235 websites in our dataset sent a string with opt-out set to ``N'' when GPC was disabled and sent a string with opt-out set to ``Y'' when GPC was enabled; this is the set of sites that sell personal information by default but that treat GPC signals as valid opt-outs in compliance with California law (denoted ``N $\rightarrow$ Y'' in Figure~\ref{fig:usps}). This number increased sharply after CPRA went into effect in January 2023 and then remained stable over the final 10 months of our study. There was no observable jump when enforcement began in July 2023. 

A small, constant group of websites sent US Privacy strings with opt-out always set to ``Y'' regardless of whether or not GPC was enabled. These are websites that claim to never sell personal information, regardless of whether or not a user has explicitly opted out. Conversely, the \texttt{N $\rightarrow$ N} category represents sites that are not opting users out of their sale even when the GPC signal is being sent; these sites are not compliant with California law. Within this category, the number of sites has decreased from 1173 to 778 during the course of our study. A small number of additional sites sent strings with less common values (e.g., ``-'', indicating that CCPA/CPRA are not applicable) or exhibited errors. 

\subsubsection{Transparency about GPC Compliance}

Finally, we investigated what information websites provide about how they respond to GPC signals. 

\begin{figure}[t]
    \centering
    \footnotesize
    \includegraphics[width=\columnwidth]{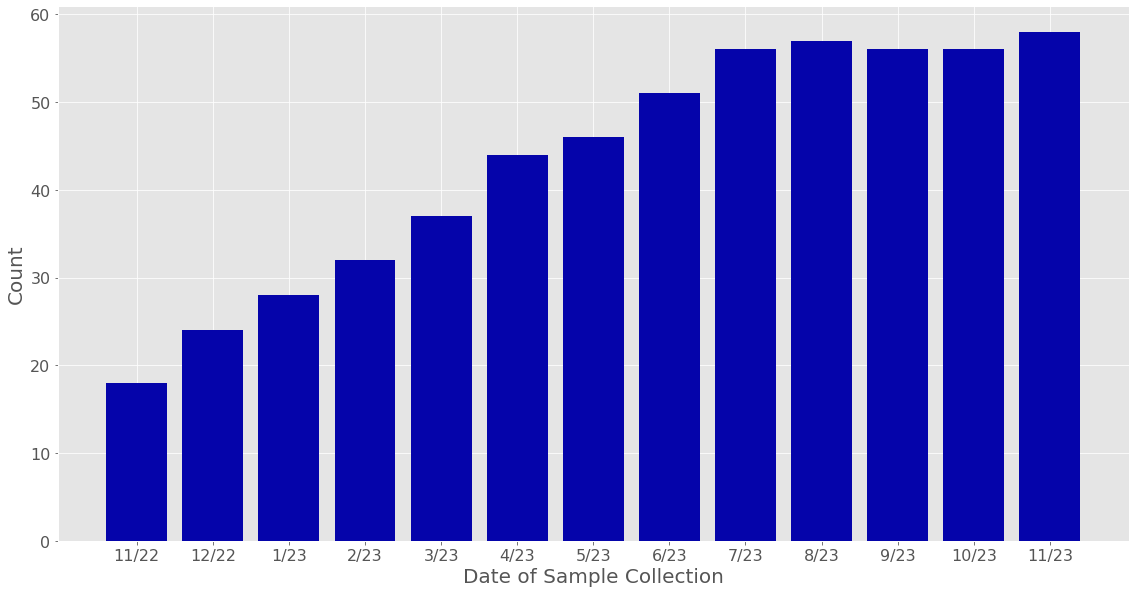}
    \caption{Number of top 25,000 websites with \texttt{GPC.json} files.}
    \label{fig:gpc_json}
\end{figure}

First, we examined whether websites provided a  \texttt{gpc.json} file in their \texttt{/.well-known} directory. According to the GPC specification, websites can indicate that they are complying with the specification by having this file and by setting the \texttt{gpc} property in this file to \texttt{true}. However, only a small number of websites have such a file.  18 websites in our dataset had a \texttt{gpc.json} file in November 2022, a number that rose steadily until enforcement of CPRA began in July 2022 (at which point 56 websites had such files) and then remained steady over the last five months of our data collection period (Figure~\ref{fig:gpc_json}).

\begin{figure}[!t]
    \centering
    \footnotesize
    \includegraphics[width=\columnwidth]{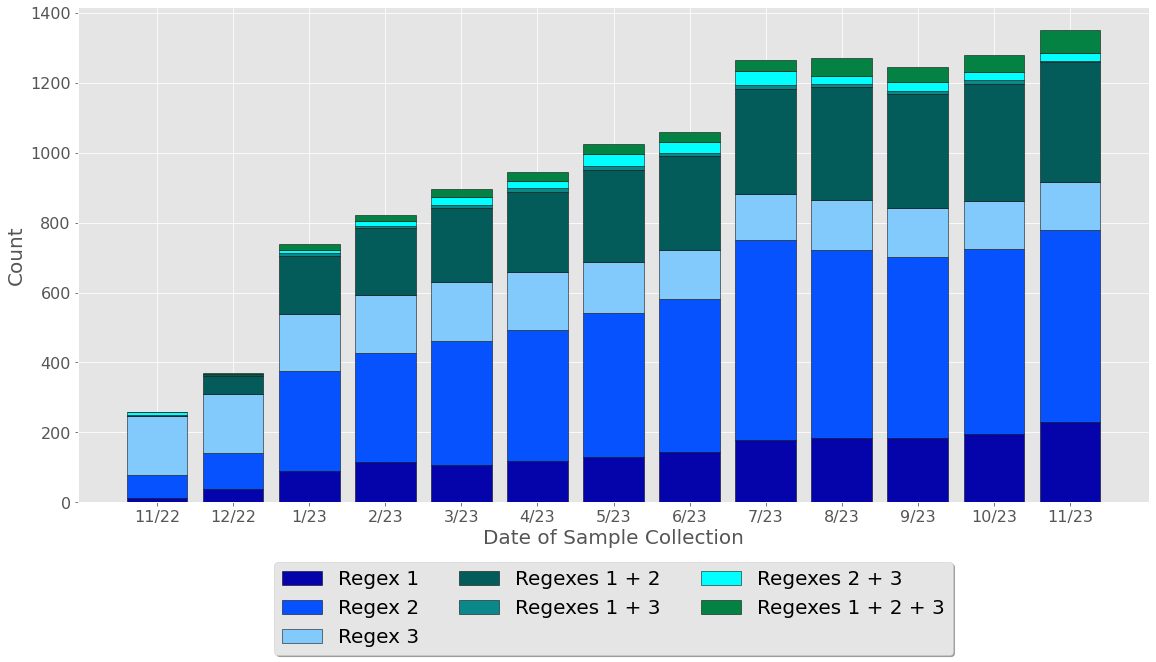}
    \caption{GPC-related expressions in privacy policies.}
    \label{fig:regex}
\end{figure}

Second, we used three regular expressions to examined the frequency with which language about GPC opt-out signals appears in privacy policies. The three expressions we used were: 
    \begin{itemize}
        \itemindent=-13pt
        \item \textbf{Regex 1:} \texttt{opt(-| )out( preference)? signals?( honored)?}
        \item \textbf{Regex 2:} \texttt{global privacy control}
	   \item \textbf{Regex 3:} \texttt{browser(-based( standard)?)? signals?}
    \end{itemize}
In November 2022, just 257 privacy policies mentioned any GPC-related expressions, with the most common terms being those relating to general browser signals (Regex 3). However, we observed jumps in the number of policies with GPC-related expressions in both January 2023 (when CPRA went into effect) and July 2023 (when enforcement of CPRA began). By November 2023, 1,350 websites in our dataset privacy policies that included one or more GPC-related expressions. Much of the increase we observed was due to websites introducing the term ``Global Privacy Control'' (Regex 2) to their privacy policy. This change is may indicate an effort to comply with the provision of CPRA that requires businesses that comply with GPC and do not provide manual opt-out links to inform users in their privacy policy about how consumers can implement an opt-out signal.

\begin{figure*}[t!]
	\centering
 	\begin{subfigure}[t]{0.495\textwidth}
		\centering
		\includegraphics[width=\linewidth]{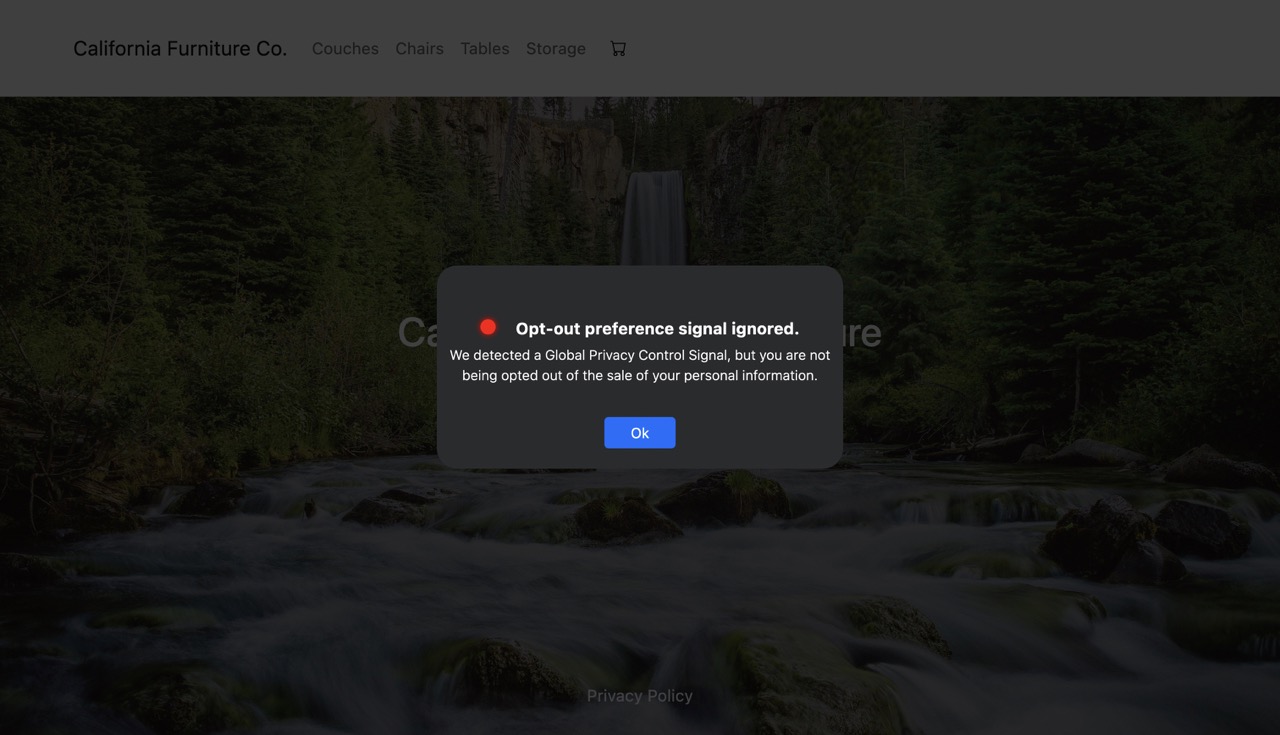} 
		\caption{Middle} \label{fig:middle}
	\end{subfigure}
    \hfill
	\begin{subfigure}[t]{0.495\textwidth}
		\centering
		\includegraphics[width=\linewidth]{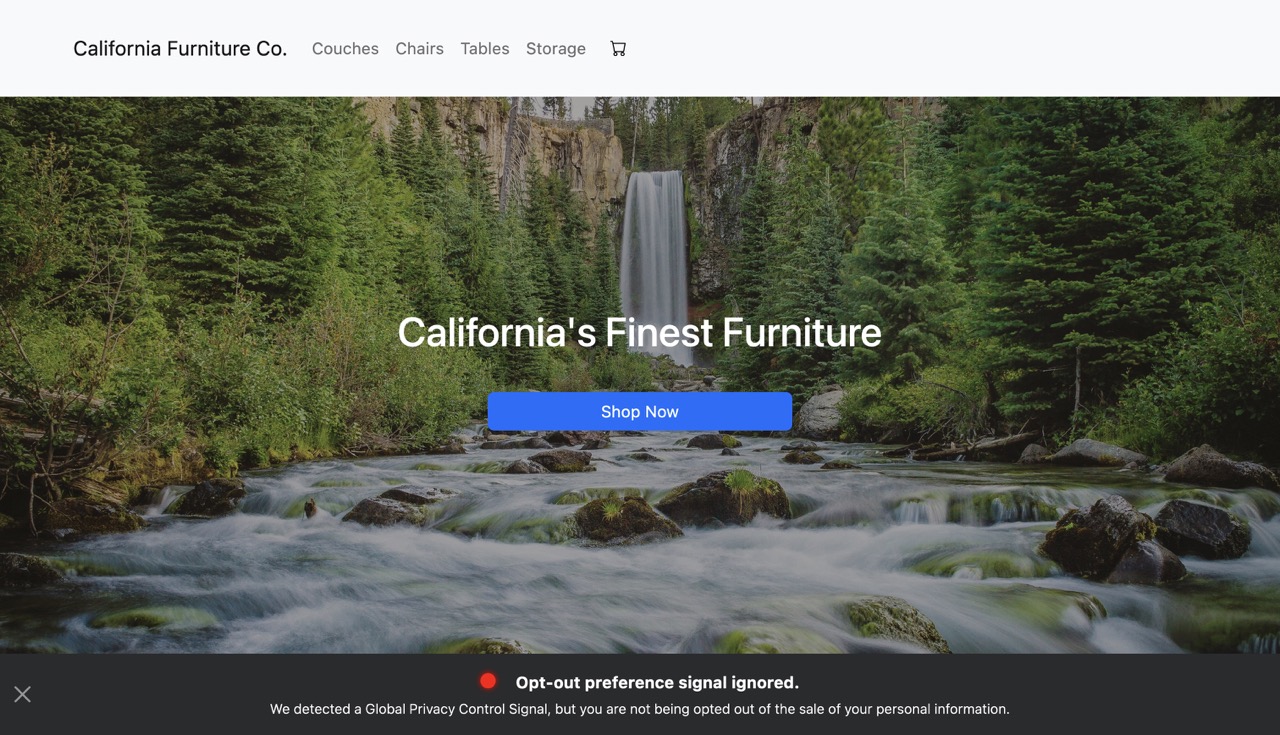} 
		\caption{Bottom} \label{fig:bottom}
	\end{subfigure}

	\vspace{1cm}
	\begin{subfigure}[t]{0.495\textwidth}
		\centering
		\includegraphics[width=\linewidth]{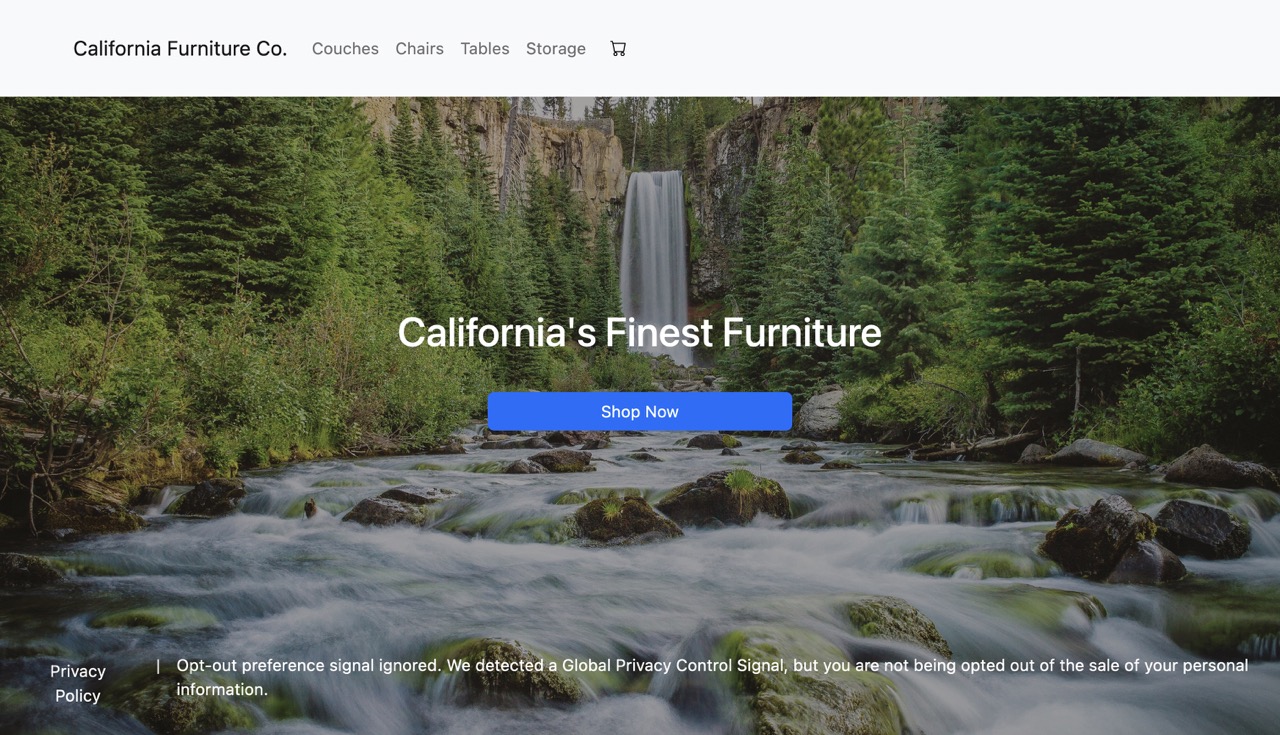} 
		\caption{Footer} \label{fig:footer}
	\end{subfigure}
	\hfill
	\begin{subfigure}[t]{0.495\textwidth}
		\centering
		\includegraphics[width=\linewidth]{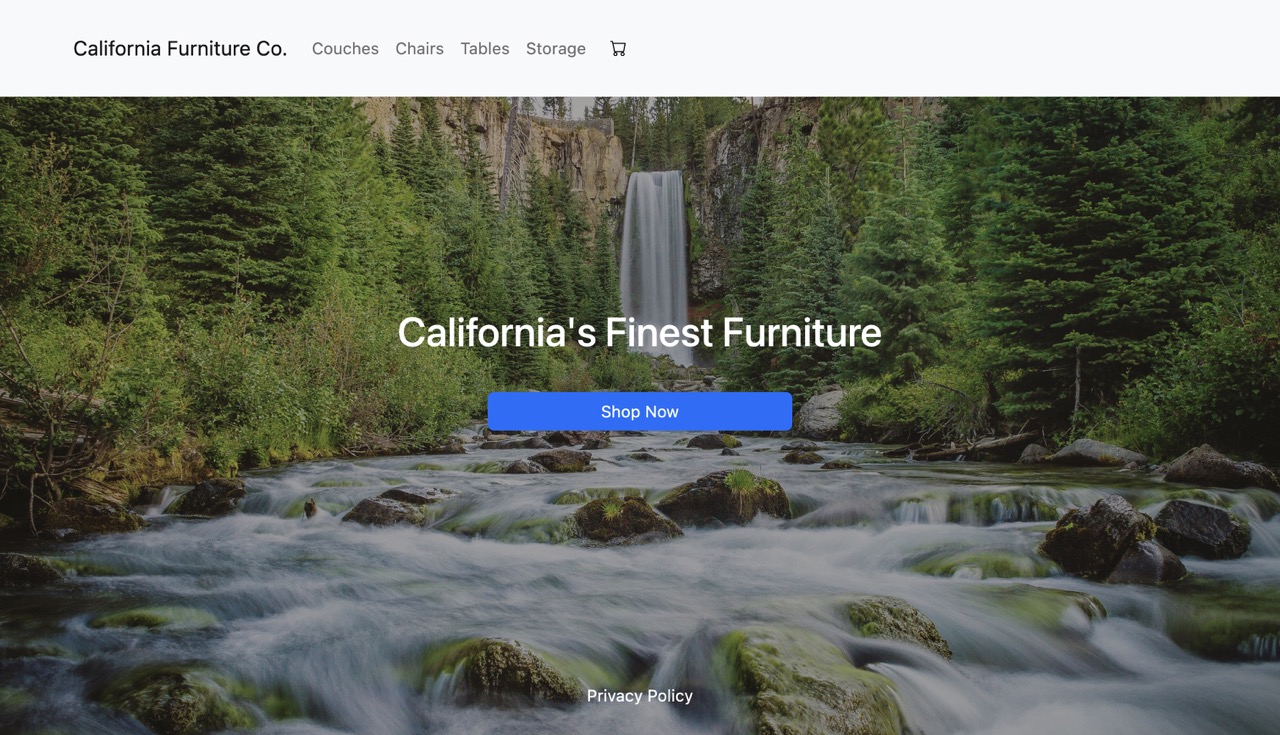} 
		\caption{None} \label{fig:none}
	\end{subfigure}
	\caption{Example of location conditions on website for ignored verbose subcondition.}
	\label{fig:example_conditions}
\end{figure*}

\section{The Impact of GPC Displays}

The third textual revision we identified that has potential to negatively impact the right to opt-out of sale under CPRA was the elimination of a requirement that websites should display whether or not they respected GPC signals; under the final CPRA guidelines, websites ``may'' provide such a notice. 
To understand the impact of this revised requirement, we conducted an online, experimental user study.

\subsection{User Study Methodology}

To conduct our study, we implemented an experimental furniture shopping website called \textit{California Furniture Company}. We elected to use a shopping website because this domain provides a credible privacy threat to a visitor: shopping habits are frequently sold to advertisement companies. Our experimental website included different conditions---display design and language---that represented various different possible regulatory regimes. 

Our conditions included four possible display designs (Figure~\ref{fig:example_conditions}):

\begin{itemize}
	\item \textbf{Middle Banner Display:} The goal of this condition was to maximize the visibility of the display information. The display was implemented as a blocking banner in the middle of the page with a three-second delay before the user could dismiss the banner. 
    The banner includes language describing whether the preference-signal was honored or ignored, along with a colored indication of GPC-compliance.
    \item \textbf{Bottom Banner Display:} This condition was designed to implement the most likely display implementation in a regulatory regime where the requirement that websites ``should'' provide a display is interpreted as requiring a banner. The banner was implemented at the bottom of the screen---the most common location for cookie banners~\cite{utz2019informed} and CCPA-related banners~\cite{o2021clear}---and provided both text and a colored indicator showing whether the website honored GPC opt-out signals. 
	\item \textbf{Footer Text-only Display:} This condition was designed to implement the most likely display implementation in a regulatory regime where information is required to appear on the homepage but no specific design elements are required. The display is implemented in the footer of the website---the most common location for CCPA opt-out of sale links~\cite{o2021clear}. The footer only includes text, but it is visible even on small screens without the participant having to scroll.
	\item \textbf{No Display:} A condition where no GPC-related display is provided. This condition is the most common implementation under the current regulatory regime. 
\end{itemize}

Participants were randomly assigned one of these four conditions. Participants assigned to a condition with a display  were then randomly assigned one of four sub-conditions to determine which text to display: 

\begin{itemize}
	\item \textbf{Honored-Verbose:} Opt-out preference signal honored. We detected a Global Privacy Control signal, so you are being opted out of the sale of your personal information.
	\item \textbf{Honored-Concise:} Opt-out preference signal honored. 
	\item \textbf{Ignored-Verbose:} Opt-out preference signal ignored. We detected a Global Privacy Control Signal, but you are not being opted out of the sale of your personal information.
	\item \textbf{Ignored-Concise:} Opt-out preference signal ignored. 
\end{itemize}

This resulted in a total of 13 conditions: 6 with displays indicating that GPC signals were honored, 6 with displays indicating that GPC signals were ignored, and 1 condition in which no information about GPC was provided (``No Display'').

\paragraph{Participant Recruitment} We recruited participants for our user study through Prolific. To avoid biasing our results by priming participants that the study was about privacy, our study was advertised as ``Online Shopping Website [BETA TEST]'' with the description ``Beta test an online shopping website and then complete a survey about the experience.'' The consent form described the study purpose and data collected in vague language that did not clearly indicate the the study was about privacy. This deception was approved in advance by the IRB at our institution, and participants were fully debriefed and provided with an opportunity to remove their data at the end of the study. 

We recruited a gender-balanced sample of participants from the United States who met two pre-screening criteria: being a resident of California and being comfortable with participating in a deception study. Participation was restricted to desktop and laptop devices. 

\paragraph{Participant Task} After consenting to participate in the study, participants were provided with a link taking them to our shopping website and the following instructions:

\begin{quote}
	The website linked below is a fake furniture shopping website that we are testing the usability of. Your task is to add a table to your cart and to check out. Once you click on check out, you will receive a completion code. 
\end{quote}

After participants had completed that task, they were asked a series of follow-up questions about their attitudes towards the website, their beliefs about the website's data practices, and their knowledge of California privacy laws and GPC.  Finally, we debriefed the participants about the deception that occurred in our study and gave them an opportunity to remove their data from the study.  

Participants who completed the task were compensated \$1.25. The median time it took our participants to complete our study was 4 minutes and 39 seconds, resulting  in an average compensation of \$16.13 per hour (slightly above California's minimum wage of \$15.50).

\paragraph{Data Collect} In addition to collecting survey responses, we also logged which web browser each participant used and whether they had GPC enabled in their browser. We also logged information about how our participants interacted with our experimental website: specifically, we recorded interactions with the opt-out mechanisms such as clicking a button on a banner as well as which pages of the website the participant visited and which links they clicked. We also logged a heartbeat whenever the webpage was in focus on the participant’s device in order to determine how long participants spent on the website. We used each participant's IP address to identify their course-grained location (e.g., country and state). All log records were associated with the participant's anonymous Prolific ID; we did not record IP addresses, fine-grained location information, or any other personally-identifiable information, and we did not link our collected data to any external databases. 

\paragraph{Ethical Considerations} Given the inclusion of human subjects in this work, we paid specific attention to ensure that our user study was designed and conducted following ethical best practices. We gave particularly careful consideration to the use of deception and to the collection of data about our human participants. 

Given the potential to influence behavior if people are aware that they are participating in a study about privacy, we decided that the use of deception would be critical to ensure data quality. To ensure that this deception was conducted ethically, we applied Prolific's ``deception'' pre-screener to our study population, ensuring that only people who had indicated they were willing to participate in deception studies would be included in our study population. We also provided a full debrief to all participants, including informing them of the option to delete their data from our dataset. We also informed the IRB at our institution that our study included deception and underwent this additional level of review and approval before conducting our study. 

We also gave careful consideration to what data we collected. We considered it important to know whether our participants were California residents, so we used IP addresses to geolocate our participants in addition to asking them where they lived. However, we were mindful of prior results about the re-identifiability of data~\cite{sweeney1997weaving} when considering what granularity of location information to record. We also took steps to ensure that IP addresses were not recorded in log entries or collected by Qualtrics, and that no other personally-identifiable information was collected. We did not force responses to demographic survey questions. 

This study was reviewed in advance and approved by the Pomona College IRB.

\begin{table}[t!]
\begin{tabular}{lccccc}
\toprule
& \multicolumn{2}{c}{Honored} && \multicolumn{2}{c}{Ignored}\\
\cline{2-3}\cline{5-6}
& Verbose & Concise && Verbose & Concise \\
\hline
Middle Banner & 49 & 45 && 38 & 41 \\
Bottom Banner & 46 & 53 && 48 & 54 \\
Footer Text & 43 & 57 && 49 & 59 \\
\hline
None & \multicolumn{5}{c}{193} \\
\bottomrule
\end{tabular}
  \caption{Breakdown of participants by condition.\label{tab:conditions}}
\end{table}

\subsection{User Study Results}

800 participants completed our study. We excluded 19 participants who self-reported as not being residents of California and 6 participants who did not enter valid task completion codes. We analyzed data collected from the remaining 775 participants.
A breakdown of our survey population by condition is provided in Figure~\ref{tab:conditions}.

Our final study population was gender-balanced and reflects the general racial breakdown of California. However, our population skews younger, more educated, and more tech-savvy than the overall population; this is consistent with prior research about the Prolific worker population~\cite{tang2022replication}. 148/775 of our participants self-reported that they had some formal education in a computer-related field,  and 163/775 participants answered that they work in a computer-related field. Detailed demographic information is provided in Table~\ref{tab:demographics}.

\begin{table}
\begin{tabular}{ll|c}
\toprule
Age & 18-24 & 153 \\
    & 24-34 & 294 \\
    & 35-44 & 162 \\
    & 45-59 & 108 \\
    & 60-75 & 51 \\
    & 75+   & 7 \\
\hline
Gender & Man & 385 \\
    & Woman & 372 \\
    & Self-describe & 7 \\
\hline
Race & While & 498 \\
    & Asian & 219 \\
    & Black/African American & 42 \\
    & Native American & 17 \\
    & Other & 53 \\
\hline 
Ethnicity & Non-Hispanic & 613 \\
    & Hispanic & 613 \\
\hline
Education & Primary School & 11 \\
    & Secondary school & 95 \\
    & Some higher education & 210 \\
    & Bachelor's degree & 358 \\
    & Graduate degree & 92 \\
    & Other & 4 \\
\bottomrule
\end{tabular}
\caption{Demographics of our user study population.}\label{tab:demographics}
\end{table}

\subsubsection{The Effect of GPC Displays on Awareness of Data Practices}

\begin{figure*}[t!]
	\centering
	\begin{subfigure}[t]{0.495\textwidth}
		\centering
		\includegraphics[width=\linewidth]{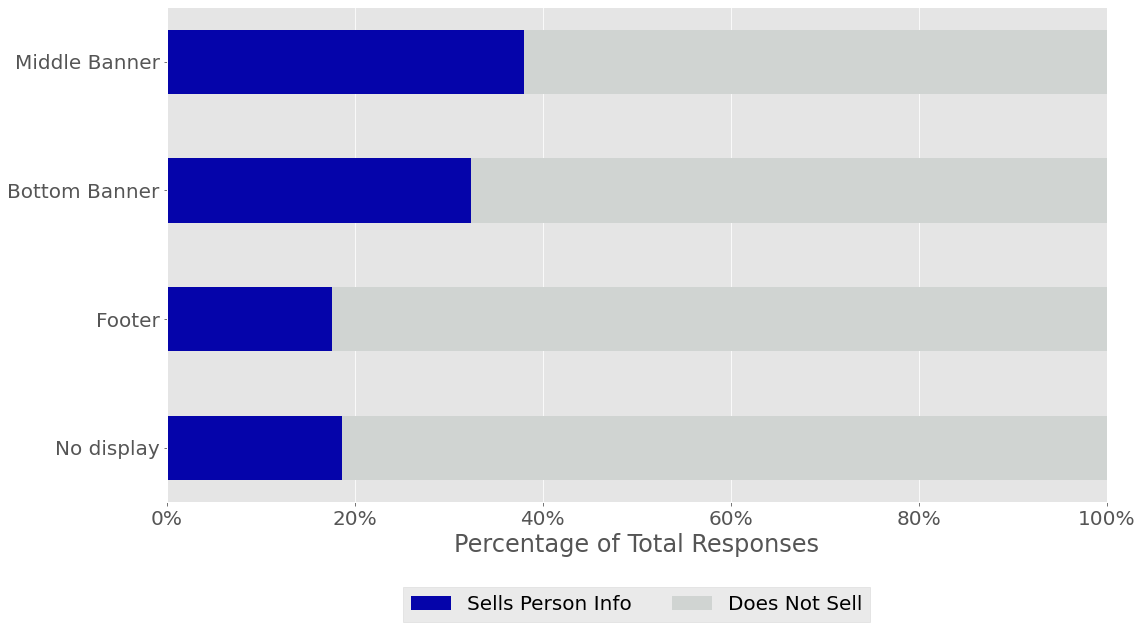} 
		\caption{When GPC signals are ignored} \label{fig:visibility-ignored}
	\end{subfigure}
	\hfill
	\begin{subfigure}[t]{0.495\textwidth}
		\centering
		\includegraphics[width=\linewidth]{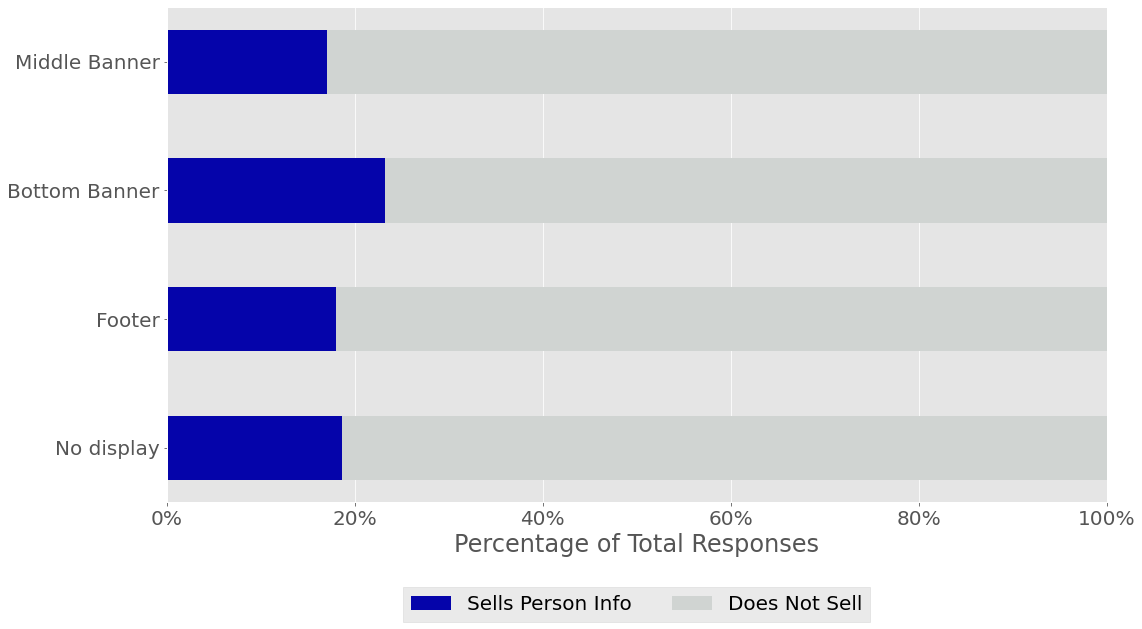} 
		\caption{When GPC signals are honored} \label{fig:visibility-honored}
	\end{subfigure}
\caption{Effect of visibility on beliefs about whether a website sells personal information.}\label{fig:visibility}
\end{figure*}

To investigate how effectively the GPC-related displays convey information to users, we asked participants whether the website sells their personal information. 

We first analyzed the effect of banner visibility on user awareness of whether a website sells their information; participant responses are summarized in Figure~\ref{fig:visibility}. For conditions that stated that GPC signals were ignored, the number of participants who ``correctly'' identified that the website sold their information increased with the visibility of the display (Figure~\ref{fig:visibility-ignored}). 38.0\% of people who saw the highly-visible middle banner recognized that the website sold their information, and 32.4\% of people who saw the bottom banner---the condition designed to reflect how websites might implement a display requirement if legal interpretation requires them to display a banner---recognized that the website sold their personal information. By contrast, only 17.6\% of participants in the footer condition identified that the website sold their information. A one-way ANOVA test found that the effect of visibility was statistically significant ($p=0.018$). For the condition with the lowest visibility---the footer text-only display---a pairwise $\chi^2$ test showed no significant difference from the condition with no display. For conditions that stated that GPC signals were honored, a one-way ANOVA test found no significant differences between the conditions with displays and the condition with no display (Figure~\ref{fig:visibility-honored}). 

\begin{figure}[!t]
	\centering
	\includegraphics[width=\linewidth]{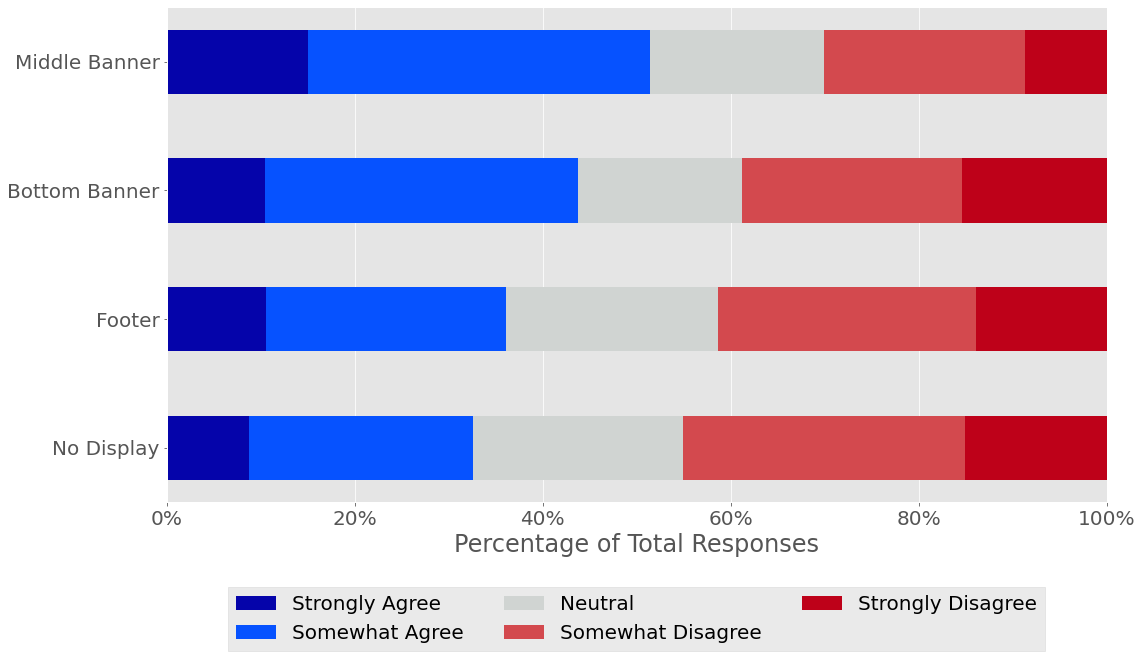} 
	\caption{Effect of visibility whether people agree with the statement: ``I am confident I am confident about my answer [about whether the website sells my personal information].''}
	\label{fig:q8}
\end{figure}

We also asked our participants how confident they were about their response to the question about whether the website sells their personal information. As expected, we found that user confidence significantly increased as the visibility of the display increased ($p=.003$), with the highest confidence for the Middle Banner condition and the lowest confidence for the condition with no display (Figure~\ref{fig:q8}). 

\begin{figure}[!t]
	\centering
	\includegraphics[width=\linewidth]{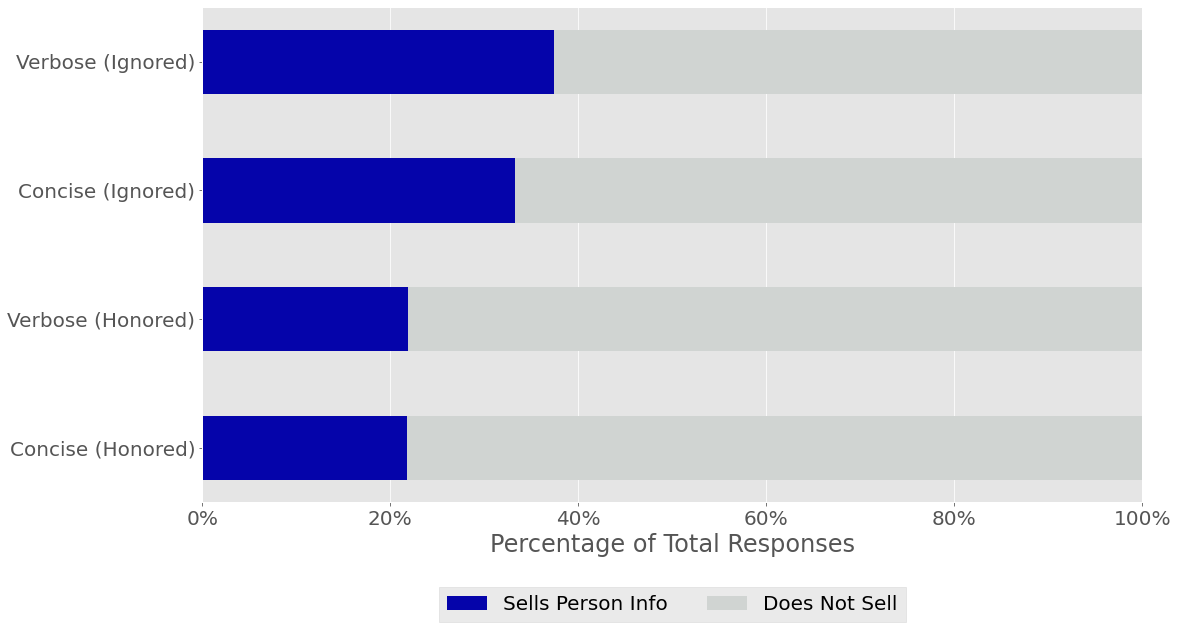} 
	\caption{Effect of display language on beliefs about whether a website sells personal information.}
	\label{fig:language}
\end{figure}

Next, we analyze the effect of language---i.e., verbose display language explaining that the user was (resp., was not) being opted out of the sale of personal information versus concise language that simply states ``Opt-out preference signal honored'' or ``Opt-out signal ignored''---on user awareness of whether a website sells their information. Since our prior results suggest that the footer display is not visible to users, we included only the middle banner and bottom banner conditions in this analysis. We found no significant differences between how many people thought a website sold their information based on language for either the conditions that ignored GPC signals or the conditions that honored GPC signals (Figure~\ref{fig:language}).

Our results suggest that by eliminating the display requirement, the final CPRA guidelines reduced user awareness of which websites continue to sell their personal information and reduced user confidence about businesses' data practices. However, the impact of that textual revision appears to depend on how a hypothetical display requirement would have been interpreted. The exact language of the display seems to have negligible impact, but the visibility of the display---which prior work relating to cookie banners suggests would depend strongly on regulatory interpretation and enforcement---does have a significant effect. If the requirement had been interpreted to allow text-only displays in the footer of a website, the impact would have been insignificant. If the requirement had been interpreted to require visible banners, it would have provided useful information to inform users when websites ignore automated opt-out signals.

\begin{figure*}[t!]
	\centering
	\begin{subfigure}[t]{0.495\textwidth}
		\centering
		\includegraphics[width=\linewidth]{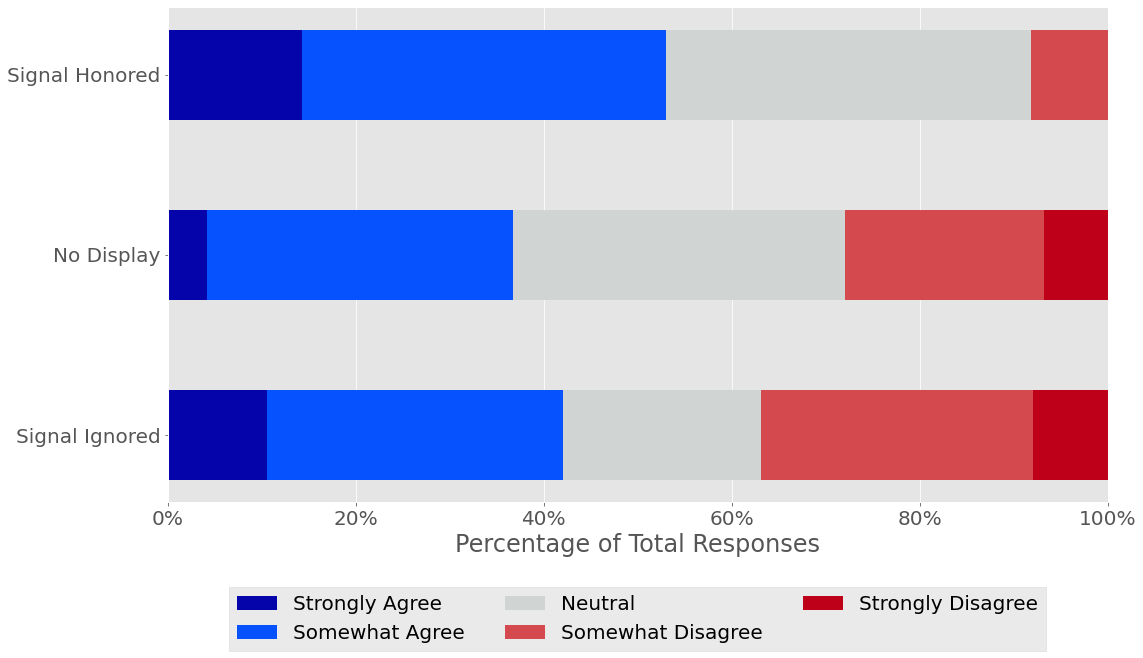} 
		\caption{``I feel like my privacy is protected on this website''} \label{fig:q2}
	\end{subfigure}
	\hfill
	\begin{subfigure}[t]{0.495\textwidth}
		\centering
		\includegraphics[width=\linewidth]{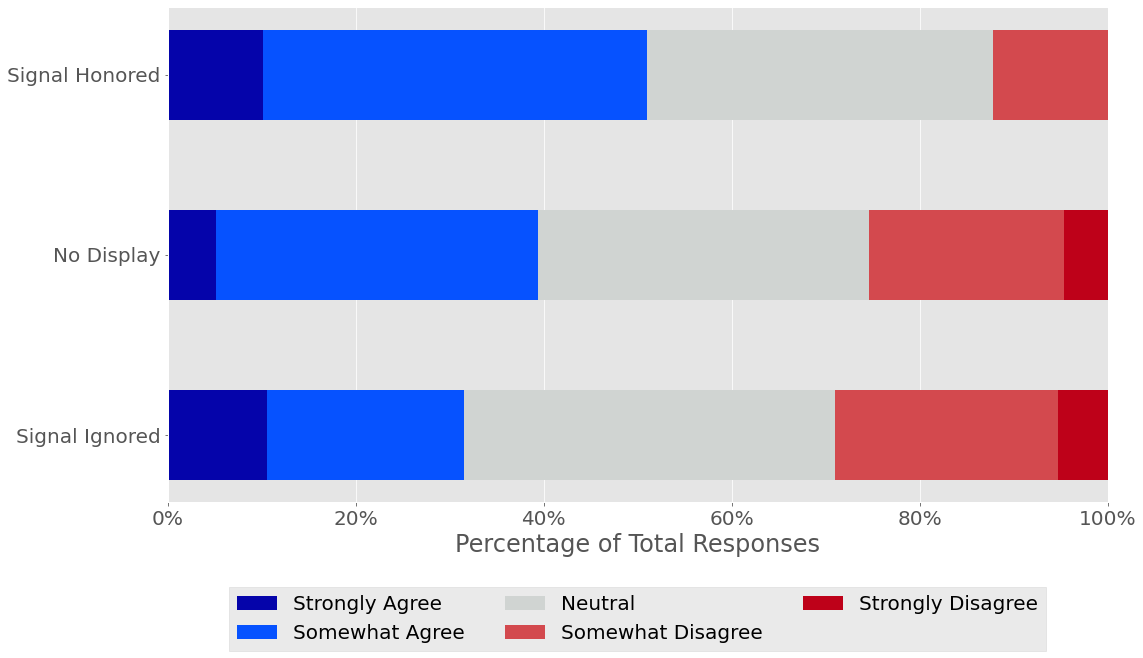} 
		\caption{"I trust this website with my personal information"} \label{fig:q1}
	\end{subfigure}
 	\hfill
	\begin{subfigure}[t]{0.495\textwidth}
		\centering
		\includegraphics[width=\linewidth]{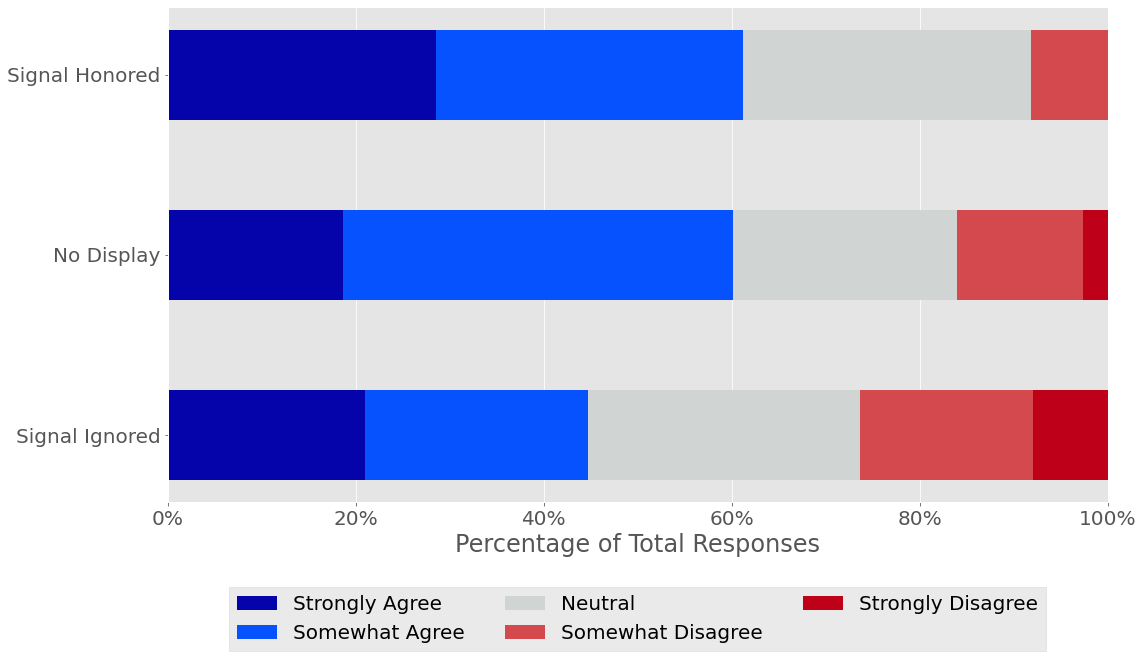} 
		\caption{"I feel at ease while I am on this website"} \label{fig:q3}
	\end{subfigure}
\caption{Effect of optimal GPC displays on user attitudes towards a website.}\label{fig:attitudes}

\end{figure*}

Our results also suggest that users implicitly assume that websites comply with GPC signals and/or do not sell personal information. That assumption does not match the actual behavior observed in our measurement study (Section~\ref{sec:measurement_study}) or website behavior observed in prior work~\cite{o2021clear}. This suggests that by eliminating a display requirement, CPRA might have missed an opportunity to enhance transparency about legal compliance and about website data practices.

\subsubsection{The Effect of GPC Displays on User Attitudes}

We asked participants a series of questions to gauge how GPC displays affected attitudes about the website. These questions asked how much participants agreed with each of three statements:
\begin{enumerate}
    \item ``I feel like my privacy was protected on this website.''
    \item ``I trust this website with my personal information.''
    \item ``I feel at ease while I am on this website.''
\end{enumerate}
We focused our analysis on comparing the condition with middle banners and verbose language to the condition with no display, since our earlier results found that this design has the highest visibility. 
We found a statistically significant difference ($p=0.004$) between the verbose middle honored, verbose middle ignored, and no display groups for the statement, ``I feel like my privacy is protected on this website''. However, there was not a large difference between the condition that where the display indicated that GPC signals were ignored and the condition with no display. We observed a similar trend for the statement ``I trust this website with my personal information'', and users reported being less at ease when GPC was ignored. However, these differences were not statistically significant. 

These results suggest that information about whether or not a website honors GPC signals---assuming it is sufficiently visible to the user---could potentially effect privacy attitudes towards a website. That in turn, suggests that the elimination of an effective display requirement for automated signals negatively impacted user privacy. However, our results also suggest that website that honor GPC signals might benefit from voluntarily displaying this compliance to users in a highly-visible way, even though such a display is not legally mandated. 

\subsubsection{User Awareness of California Privacy Rights and Mechanisms}

While the primary goal of this work was to understand the impact of textual revisions on user privacy, ours is the first large-scale user study about California privacy laws to be conducted since CPRA went into effect. We therefore also took the opportunity to quantify Californians' awareness of their rights under CCPA and their experience with various mechanisms for invoking their right to opt-out of sale. 

\begin{figure}[!t]
	\centering
	\includegraphics[width=\linewidth]{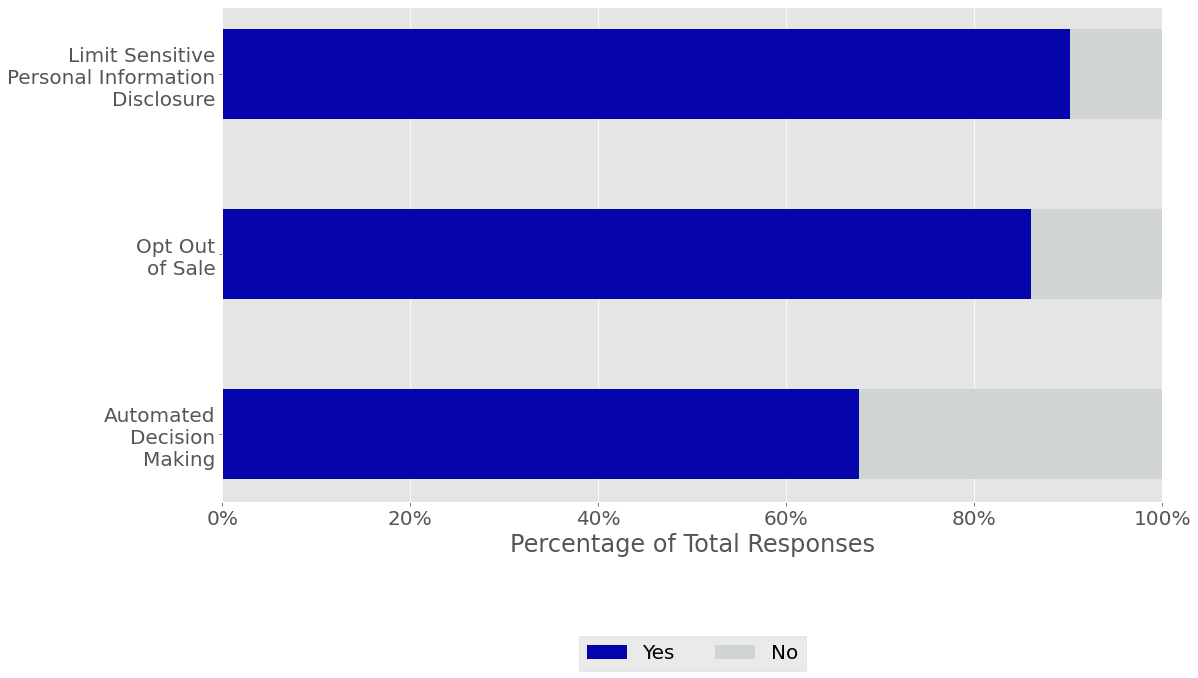} 
	\caption{Responses to questions about familiarity with California privacy law.}
	\label{fig:q9-11-13}
\end{figure}
To understand Californians' awareness of their privacy rights, we asked each study participant whether they believe various rights exist under current California law. We asked about three rights: 
\begin{enumerate}
    \item A right to limit the sharing of sensitive personal information.
    \item A right to opt-out of sale of personal information
    \item A right to opt-out of automated decision making. 
\end{enumerate}
The results of these survey questions are displayed in Figure~\ref{fig:q9-11-13}. 

The first right we asked about---limiting sharing of sensitive personal information---is a new right introduced by CPRA. Interestingly, 90.2\% of participants correctly responded that they had such a right. However, this might indicate that many people previously (incorrectly) assumed that they had such a right under CCPA rather than indicating that people are aware that such a right was introduced in 2023. 86.1\% of our participants correctly identified that they have a right to opt-out of sale---a right introduced by CCPA in 2020 and maintained by CPRA. This represents a significant increase from the 58.9\% who were aware of this right  in 2020~\cite{o2021clear}, suggesting that awareness of this right has increased in the three years since CCPA went into effect. However, 67.7\% of our participants incorrectly believed that they have a right to opt-out of automated decision making; no such right exists (or has ever existed) under California law. This suggests that people may be generally aware that they have privacy rights under California law without being aware of exactly what these rights are, potential resulting in overly-optimistic assumptions about what rights they currently have.

\begin{figure}[!t]
	\centering
	\includegraphics[width=\linewidth]{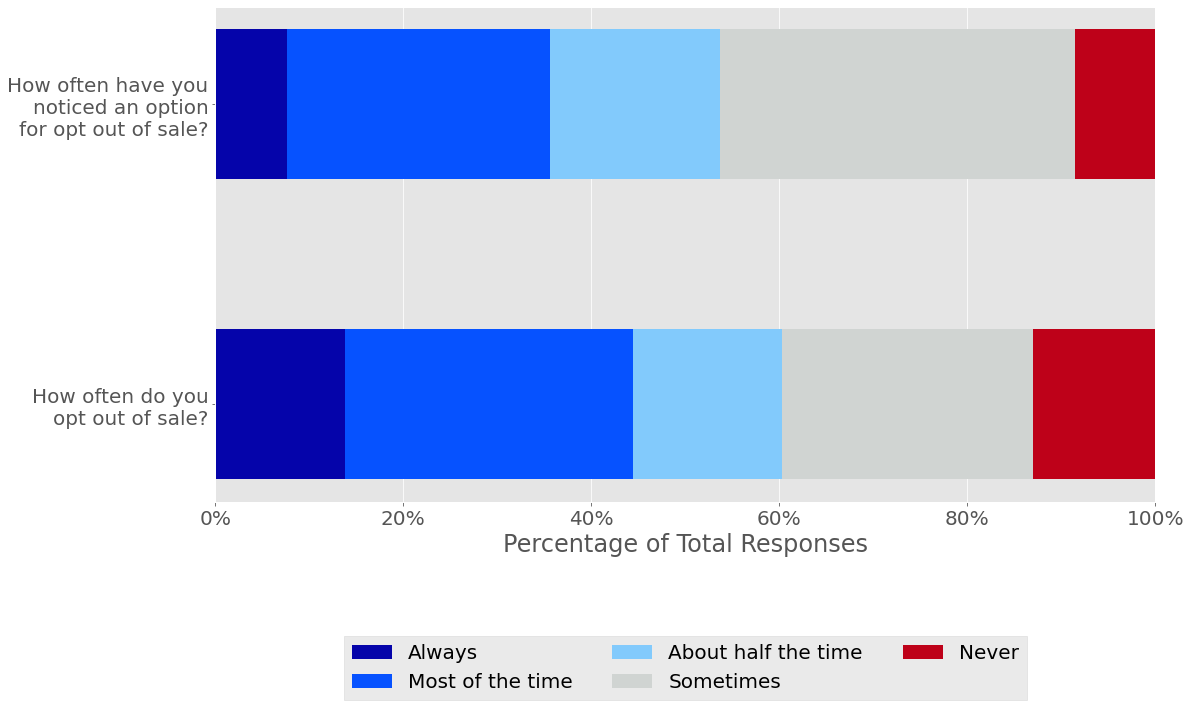} 
	\caption{Responses to questions about opt out of sale options.}
	\label{fig:q15-16}
\end{figure}

After asking about users' awareness of California privacy rights, we confirmed to each participant that California does in fact grant a legal right to opt-out of sale of personal information.  We then asked how often they have noticed such options and how often they invoke their right to opt-out of sale. Over 30\% of participants claimed to notice opt-out of sale notices always or most of the time and approximately 40\% claimed to opt out of sale always or most of the time (Figure~\ref{fig:q15-16}).

\begin{figure}[!t]
	\centering
	\includegraphics[width=\linewidth]{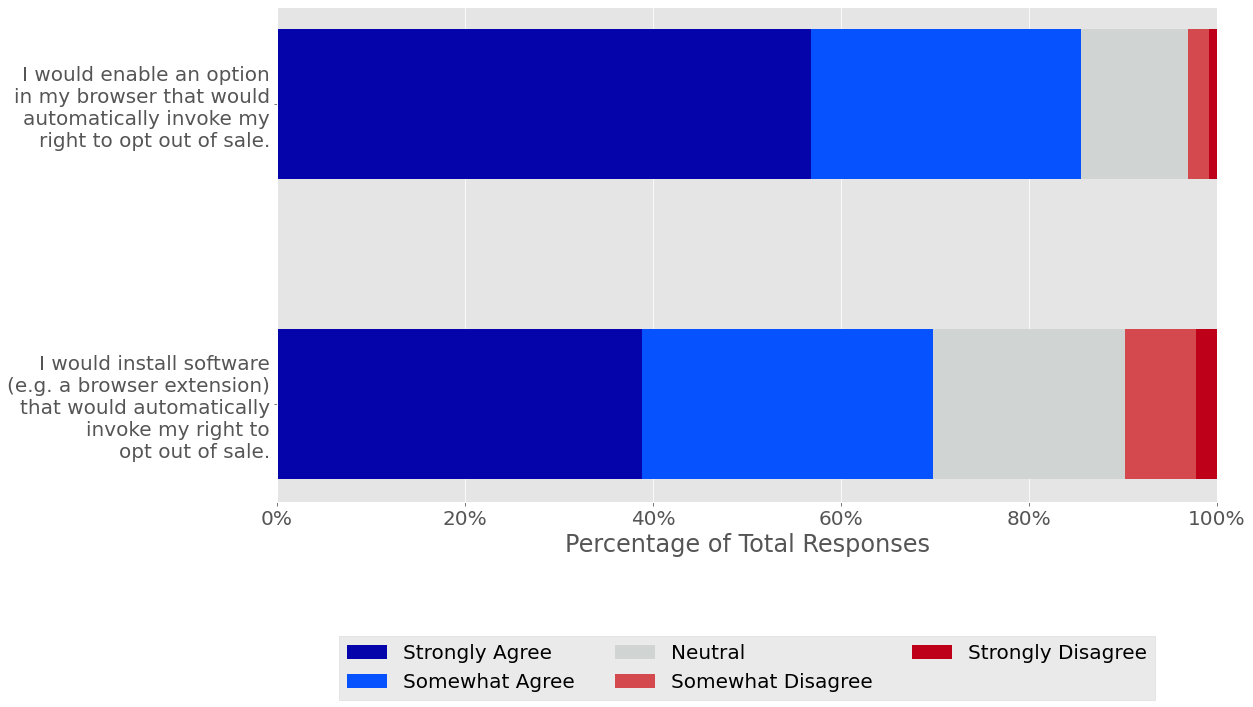} 
	\caption{Responses to questions about enabling option in browser and installing software to opt out of sale.}
	\label{fig:q19-20}
\end{figure}
We also explored user awareness of GPC. We asked each user to briefly describe GPC and qualitatively coded the responses. Only 5.9\% of responses indicated an understanding of what GPC is. Most participants simply responded that they didn't know. We also observed some misconceptions, such as a belief that GPC notifies users when websites are selling their information (``a web browser setting that notifies users if their information is being sold and gives the ability to opt out'') or a belief that GPC is a general privacy signal with broad impact (``Global Privacy Control is a web standard that allows users to exercise their privacy rights and opt-out of online tracking.''). 9.7\% of our participants completed the study from a browser that had GPC enabled; however, only 5 of these 75 participants were able to correctly identify that they had GPC enabled in their browser (8 said they did not and 62 did not know). This suggests that GPC is not yet broadly understood, and that a significant number of users who currently have GPC enabled may not be aware of their settings or be aware of of what exactly GPC is. 

Finally, we asked whether participants would enable a browser setting that invoked their right to opt-out of sale and whether they would install software that would invoke this right. 85.5\% agreed that they would enable a browser setting, while a significantly lower number (69.8\%) claimed that they would install software. These results further emphasize  the importance of built-in browser support rather than relying on extensions or other software. Currently, GPC is not supported at all on Safari, and it can only be enabled on Chrome or Edge after installing a browser extension. Browsers that have a setting to enable GPC---such as Firefox and Brave---represent less than 5\% of the U.S. marketshare. 
\section{Discussion}

Our results provide a nuanced picture of California Internet privacy under CPRA. While many provisions of CPRA expand privacy rights---such as the introduction of a right to limit the sharing of sensitive personal information, the expansion of the right to delete to cover third parties, the codification of automated opt-out signals, and the definition of dark patterns that preclude consent---CPRA also introduced textual revisions that negatively impact privacy. Our results also have potential implications for businesses, for regulators, and for future research in this space.

\begin{rec}
Lawmakers and regulators should write and interpret regulations to guarantee both manual and automated options for invoking privacy rights.
\end{rec}

Prior work has clearly demonstrated limitations of manual opt-out mechanisms that implemented the right to opt-out of sale under CCPA~\cite{o2021clear,3websitestudy}, so the interpretation that companies must accept automated opt-out signals has potential to signification ease the burden placed on users who want to invoke their right to opt-out of sale. Our results suggest that users like the idea of an automated opt-out signal and that many users are interested in enabling such a signal. However, we also found that manual opt-out links had higher visibility and higher use (albeit usually not consistently) among our study population. These results, together with the reality that automated signals are not consistently available or accessible on all platforms---suggest that lawmakers and regulators should incorporate automated signals into privacy regimes as a complement to, rather than a replacement for, manual opt-out mechanisms.

\begin{rec}
Lawmakers and regulators should write and interpret regulations to require companies to clearly identify whether they are subject to particular legal regulations. 
\end{rec}

A challenge in this work was identifying which websites were required to implement the right to opt-out of sale.  We side-stepped this challenge by focusing primarily on overall trends relating to the right to opt-out of sale. We also focused parts of our analysis on the set of 2,429 websites that complied with CCPA in November 2022, a set that was relatively easy to identify because CCPA required standardized language for opt-out of sale links on all websites subject to that law and by recording U.S. Privacy strings both with and without GPC enabled. However, the textual revisions introduced by CPRA now allow websites to legally omit this link, and U.S. Privacy Strings will be deprecated in January 2024. Identifying which websites and business are subject to---or complying with---California law is therefore likely to become increasingly challenging. 
Regulators and future lawmakers should issue new guidelines that require standardized displays of whether or not privacy laws apply and should require such displays in future regulations.
In the meantime, computer scientists should develop new, scalable heuristics for identifying which companies are subject to various regulations. 

\begin{rec}
Browsers---including both desktop and mobile versions---and other Internet-accessible devices should consistently support automated opt-out signals.
\end{rec}

While we find that many users would be interested in enabling GPC, this signal is not yet supported natively in most browsers. Firefox and Brave have a preferrence setting that allows user to enable GPC. However, the current browsers with the most marketshare do not. There are Chrome extensions available that add support for GPC, but the must be installed separately, something that our results show users are less willing or likely to do. Safari currently provides not support for GPC. Browser vendors and makers of other IoT or Internet-connected devices should ensure that there is consistent, built-in support for sending automated opt-out signals. Preliminary work suggest that mechanism design for enabling opt-out signals significantly affects user decisions~\cite{zimmeck2024generalizable}; browser vendors should further investigate best practices for empower users to enable opt-out signals that accurately reflect privacy preferences. 

\begin{rec}
Websites should clearly communicate whether they honor GPC signals.
\end{rec}

Our results suggest that user assumptions about website behavior---particularly how websites respond to opt-out signals---is not always consistent with empirical observations, and that a clear statement that websites honor GPC would both make users feel more protected and enhance transparency. It could also facilitate legal enforcement. Moreover, our results provide preliminary evidence that visible displays communicating that GPC signals are honored might improve user attitudes towards a website. This could motivate websites that honor GPC to voluntarily adopt visible displays.

\begin{rec}
Researchers working at the intersection of computer science and privacy law should conduct longitudinal studies that continue past the date when a law goes into effect. 
\end{rec}

CPRA codified the requirement that websites must treat automated signals like GPC as valid opt-outs under California law. 
However, the requirement that websites treat GPC as opt-out signals pre-dates CPRA. The California Attorney General announced that his office would be interpreting CCPA as requiring GPC compliance in January 2021~\cite{gpctweet,cppafaqs}, and that office levied the first significant fine for ignoring GPC signals against Sephora in August 2022~\cite{violation}. Post-enactment guidelines and case law are universally acknowledged in the legal community as providing essential interpretations of laws that define the rights and obligations imposed by those laws. However, longitudinal studies conducted by the computer science community to date have focused on the periods shortly before and after a law goes into effect. To provide meaningful evaluations of  privacy regulations, future work will need to conduct longitudinal measurement studies over longer period after a law goes into effect, specifically including the evaluation of post-enactment events that change the legal interpretation of privacy rights. 



\section{Conclusion}

This work provides the first quantitative evaluation of the right to opt-out of sale under CPRA. 
We find that although the number of sites that honor opt-out signals has increased, the number of sites that only accept automated signals and provide no manual opt-out mechanism has also increased. Since many browsers to not yet support automated signals, this reduced set of available mechanisms reduces accessibility of the right to opt-out of sale. 
We also find that many websites that previously provided opt-out of sale links no longer provide any mechanism to opt-out of sale; this confirms that the reduced scope of CPRA has also negatively impacted privacy. Our user study shows that the requirement to display how opt-out signals are processed---which was omitted from the final language of the CPRA guidelines---represents a missed opportunity to enhance transparency and improve awareness of Internet privacy rights under CPRA. These results provide a nuanced picture of the current state of the right to opt-out of sale under CPRA.


\begin{acks}
This work was supported by in part by NSF grant 2317115 and by internal funds from Pomona College. 
\end{acks}

\newpage
\bibliographystyle{ACM-Reference-Format}
\bibliography{refs}

\appendix
\section{User Study Survey Questions}

The following questions were asked to our participants after they interacted with the website:

\footnotesize
\begin{enumerate}
    \item[(1)] "Please rate how much you agree with the following statement. I trust this website with my personal information." (Strongly agree / Somewhat agree / Neutral / Somewhat disagree / Strongly disagree)
    \item[(2)] "Please rate how much you agree with the following statement. I feel like my privacy is protected on this website." (Strongly agree / Somewhat agree / Neutral / Somewhat disagree / Strongly disagree)
    \item[(3)] "Please rate how much you agree with the following statement. I feel at ease while I am on this website." (Strongly agree / Somewhat agree / Neutral / Somewhat disagree / Strongly disagree)
    \item[(4)] "Please rate how much you agree with the following statement. I would visit this website again." (Strongly agree / Somewhat agree / Neutral / Somewhat disagree / Strongly disagree)
    \item[(5)] "Please rate how much you agree with the following statement. I would be likely to use this website." (Strongly agree / Somewhat agree / Neutral / Somewhat disagree / Strongly disagree)
    \item[(6)] "How much do you agree with the following statement: I am comfortable with websites selling my personal information to third party companies?" (Strongly agree / Somewhat agree / Neutral / Somewhat disagree / Strongly disagree)
    \item[(7)] "From what you observed on this website, does this website sell your personal information?" (Yes / No)
    \item[(8)] "How much do you agree with the following statement: I am confident about my answer to the previous question." (Strongly agree / Somewhat agree / Neutral / Somewhat disagree / Strongly disagree) 
    \item[(9)] "To the best of your knowledge, does California law give users the right to limit the disclosure of sensitive personal information?" (Yes / No)
    \item[(10)] "How much do you agree with the following statement: I am confident about my answer to the previous question." (Strongly agree / Somewhat agree / Neutral / Somewhat disagree / Strongly disagree)
    \item[(11)] "To the best of your knowledge, does California law require that websites that sell your data allow you to opt out of the sale of your personal information?" (Yes / No)
    \item[(12)] "How much do you agree with the following statement: I am confident about my answer to the previous question." (Strongly agree / Somewhat agree / Neutral / Somewhat disagree / Strongly disagree)
    \item[(13)] "To the best of your knowledge, does California law require that websites allow individuals to opt out of automated decision making?" (Yes / No)
    \item[(14)] "How much do you agree with the following statement: I am confident about my answer to the previous question." (Strongly agree / Somewhat agree / Neutral / Somewhat disagree / Strongly disagree)
    \item[(15)] "Under California law, websites are legally obligated to give you an option to opt out of the sale of your personal information. How often have you noticed websites you visit giving you such an option?" (Never / Sometimes / About half the time / Most of the time / Always)
    \item[(16)] "How often do you opt out of the sale of your personal information on websites you visit?" (Never / Sometimes / About half the time / Most of the time / Always)
    \item[(17)] "In 20 words or less, what is Global Privacy Control? If you are don't know, please just say \textit{I don't know}." (Free response)
    \item[(18)] "Do you currently have Global Privacy Control (GPC) enabled?" (Yes / No / I don't know)
    \item[(19)] "How much do you agree with the following statement: If it were available, I would enable an option in my browser that would automatically invoke my right to opt out of sale on all websites I visit." (Strongly agree / Somewhat agree / Neutral / Somewhat disagree / Strongly disagree) 
    \item[(20)] "How much do you agree with the following statement: If it were available, I would install software (e.g. a browser extension) that would automatically invoke my right to opt out of sale on all websites I visit." (Strongly agree / Somewhat agree / Neutral / Somewhat disagree / Strongly disagree) 
    \item[(21)] "What is your current age?" (18-24 / 25-34 / 35-44 / 45-59 / 60-74 / 75+)
    \item[(22)] "What is your gender?" (Man / Woman / Prefer not disclose / Prefer to self describe: \underline{\hspace{2cm}})
    \item[(23)] "Choose one or more races that you consider yourself to be." (White / Black or African American / American Indian or Alaska Native / Asian / Pacific Islander or Native Hawaiian / Other)
    \item[(24)] "Do you consider yourself to be Hispanic?" (Yes / No)
    \item[(25)] "In which state do you currently reside?" (50 states / American Samoa / District of Columbia / Guam / Minor Outlying Islands / Northern Mariana Islands / U.S. Virgin Islands / Not in U.S.)
\end{enumerate}

\end{document}